\documentclass[fleqn,usenatbib]{mnras}

\usepackage{newtxtext,newtxmath}

\usepackage[T1]{fontenc}

\usepackage{graphicx}
\usepackage{amsmath}
\usepackage[version=4]{mhchem}

\title[Modeling the Thermal Structure of a Protoplanetary Disk]
{Modeling the Thermal Structure of a Protoplanetary Disk Using
Multiband Flux-Limited Diffusion Approximation}
\author[Pavlyuchenkov \& Akimkin]{Ya. N. Pavlyuchenkov*, V. V. Akimkin  \\
        Institute of Astronomy of the Russian Academy of Sciences, Moscow, 119017 Russia}
\date{}
\begin{document}

\maketitle

\begin{abstract}
{
This work continues the analysis of the model for calculating the thermal
structure of an axisymmetric protoplanetary disk, initiated in the
paper by~\cite{2024ARep...68.1045P}. The model is based on the well-known
Flux-Limited Diffusion (FLD) approximation with separate calculation of
heating by direct stellar radiation (hereinafter referred to as the
FLD$^{\rm s}$ method). In addition to the previously described FLD$^{\rm
s}$ model with wavelength-averaged opacities, we present a multiband
model mFLD$^{\rm s}$, where the spectrum of thermal radiation is divided
into several frequency bands. The model is based on an implicit
finite-difference scheme for the equations of thermal radiation
diffusion, which reduces to a system of linear algebraic equations
written in hypermatrix form. A modified Gauss method for inverting the
sparse hypermatrix of the original system of linear equations is
proposed. The simulation results described in the article show that the
midplane radial temperature profile obtained with the mFLD$^{\rm s}$ method
has a variable slope in accordance with the reference Monte Carlo radiative transfer
simulations. The mFLD$^{\rm s}$ model also qualitatively reproduces the
non-isothermality of the temperature distribution along the angular
coordinate near the midplane, which is not provided by the
FLD$^{\rm s}$ method. However, quantitative differences remain between
the reference temperature values and the results of mFLD$^{\rm s}$. These
differences are likely due to the diffusive nature of the FLD
approximation. It is also shown that the characteristic times for the
disk to reach thermal equilibrium within the mFLD$^{\rm s}$ model can be
significantly shorter than in FLD$^{\rm s}$. This property should be taken
into account when modeling non-stationary processes in protoplanetary
disks within FLD-based models.}
\vspace{0.3cm}
\end{abstract}

\maketitle

\section{Introduction}

Star formation is accompanied by the formation of circumstellar
protoplanetary disks, the evolution of which ultimately leads to the
emergence of planets. The physical processes governing the evolution of
protoplanetary disks are extremely diverse, and many of them depend on
the thermal structure of the disk. The temperature distribution in the
disk also affects its observable properties in the infrared range. The
rates of heating and cooling processes determine the development of a
number of instabilities in protoplanetary disks: gravitational,
convective, thermal, vertical shear instability, baroclinic, pulsational
(convective overstability), and others, see, for example, reviews
by~\cite{2015arXiv150906382A,2023ASPC..534..465L}.

Due to the variety of processes and the mathematical complexity of the
problem, the calculation of the self-consistent temperature structure of
the disk and its thermal evolution is generally carried out using
numerical modeling. Significant progress has been made in the development
and use of numerical methods for modeling radiation transfer in
astrophysical problems in general and in application to circumstellar
disks in particular, see e.g.~\cite{2019FrASS...6...51T, 2024FrASS..1146812W}.
However, many aspects of the methods used and their areas of application
have not been fully analyzed. The choice of a method for calculating
thermal evolution in dynamic problems, where not only accuracy but also
speed is important, remains a relevant task.

In the work by~\cite{2024ARep...68.1045P} (hereinafter referred to as
Paper~I), the FLD$^{\rm s}$ model for calculating the non-stationary
thermal structure of a protoplanetary disk in an axisymmetric
approximation was illustrated in detail. It is based on the widely used
approach of separating the radiation field into stellar and intrinsic
thermal radiation of the medium. Heating by stellar radiation was
calculated using a ray-tracing method, and the well-known Flux-Limited
Diffusion (FLD) approximation was used to describe thermal radiation
\citep{1981ApJ...248..321L}. A comparison of the calculation of the
stationary thermal structure of a protoplanetary disk within FLD$^{\rm
s}$ with more accurate calculations based on the accelerated
$\Lambda$-iteration method~\citep{2012MNRAS.421.2430P}
showed that the FLD$^{\rm s}$ method excellently reproduces the
temperature in the upper and near-surface layers of the disk, but in the
disk midplane, it may differ from the exact solution, both
in the slope of the radial profile and in its normalization. The
temperature distribution from FLD$^{\rm s}$ in the disk's interior (i.e.,
in the region optically thick to stellar radiation) turned out to be
close to isothermal in the vertical direction, which also does not agree
with the results of the exact calculation. In Paper~I, it was suggested
that these differences are related to the diffusive nature of the FLD
approximation, but the aspect related to the use of wavelength-integrated
FLD equations was not analyzed. Indeed, in the implemented FLD$^{\rm s}$
method, the values of the radiation energy density $E$, averaged over the
entire spectrum, as well as the spectrum-averaged absorption coefficients
(so-called Planck and Rosseland opacities), were used to describe thermal
radiation. Meanwhile, in the work by~\cite{2002A&A...395..853D}, within the
framework of a 1+1D model of a protoplanetary disk, it was shown that the
use of spectrum-averaged opacities leads to an isothermal temperature
distribution in the vertical direction near the midplane. The need to move
to a more accurate approximation led to the development of methods based
on FLD or M1 closure theory, in which the spectrum of thermal radiation
is divided into several intervals, see, for
example,~\cite{2011ApJS..194...23V,2012A&A...543A..60V}. However, the use
of such methods for modeling the evolution of protoplanetary disks is
still rare and requires more thorough analysis.

In this article, we continue the analysis of the FLD method for modeling
the thermal structure of a protoplanetary disk by implementing its
multiband version, which we will henceforth refer to as mFLD$^{\rm s}$.

\section{Method Description}

\subsection{Non-stationary Equations of Thermal Evolution and Radiation Diffusion}

Before proceeding to the description of the mFLD$^{\rm s}$ method, let us
recall the main equations of the basic FLD$^{\rm s}$ method from Paper~I:
\begin{eqnarray}
\rho c_{\rm V} \frac{\partial T}{\partial t}&=& c \alpha \left(E - a T^4\right) + S
\label{therm_sys1}\\
\frac{\partial E}{\partial t}&=& - c \alpha \left(E - a T^4\right) + \hat{\Lambda} E,
\label{therm_sys2}
\end{eqnarray}
where $\rho$ is the density of the gas-dust medium, $c_{\rm V}$ is the
specific heat capacity of the medium [erg K$^{-1}$ g$^{-1}$], $c$ is the
speed of light, $\alpha$ [cm$^{-1}$] is the spectrum-averaged absorption
coefficient of thermal radiation (excluding scattering, per unit volume
of the gas-dust medium), $a$ is the radiation density constant,
$S$ [erg cm$^{-3}$ s$^{-1}$] is the heating rate
by stellar radiation, $T$ is the temperature of the medium, and $E$ is
the energy density of thermal radiation.

Equation~\eqref{therm_sys1} describes the change in the thermal energy density
of the medium as a result of the absorption and
re-emission of thermal radiation (terms $c\alpha E$ and $c\alpha a T^4$,
respectively), as well as due to the absorption of direct stellar
radiation, determined by the function $S$. Equation \eqref{therm_sys2} is
a moment equation of radiation transfer and describes the change in the
energy density of radiation as a result of the absorption and re-emission
of thermal radiation, as well as due to the spatial diffusion of thermal
radiation, represented by the operator $\hat{\Lambda}E$. It is assumed
that the main heat capacity of the medium is due to the gas, and the
opacity is due to dust. It is also assumed that heat exchange between gas
and dust is efficient enough to maintain their temperatures equal, which
is well satisfied for the bulk of the matter in the protoplanetary disk
but may break down in the rarefied atmosphere of the disk.

The idea of the mFLD$^{\rm s}$ method is to split the spectral range into
intervals, introducing the radiation energy density, absorption
coefficients, and emission coefficients within each interval. The
corresponding system of equations takes the form:
\begin{eqnarray}
\rho c_{\rm V} \frac{\partial T}{\partial t}&=& c \sum_{m=1}^{M}\left[\alpha_m(T) E_m - \epsilon_m(T)\right] + S,
\label{therm_sys1b} \\
\frac{\partial E_m}{\partial t}&=& - c\left[\alpha_m(T) E_m - \epsilon_m(T)\right] + \hat{\Lambda}_m E_m,
\label{therm_sys2b}
\end{eqnarray}
where the index $m$ indicates that the quantity belongs to the $m$-th frequency interval,
and $M$ is the total number of frequency intervals.
The physical quantities appearing in the
system~\eqref{therm_sys1b}--\eqref{therm_sys2b} are defined as follows:
\begin{eqnarray}
\alpha_m(T) &=& {\int\limits_{\nu_m}^{\nu_{m+1}}\alpha_{\nu}B_{\nu}(T)d\nu}\bigg/
{\int\limits_{\nu_m}^{\nu_{m+1}}B_{\nu}(T)d\nu}, \\
\epsilon_m(T) &=& \dfrac{4\pi}{c} \int\limits_{\nu_m}^{\nu_{m+1}}\alpha_{\nu}B_{\nu}(T)d\nu,
\end{eqnarray}
where $\alpha_{\nu}=\rho\kappa_{\nu}$ is the spectral absorption
coefficient [cm$^{-1}$], $B_{\nu}(T)$ is the intensity of blackbody
radiation, and $\nu_m$ and $\nu_{m+1}$ correspond to the boundaries of
the $m$-th frequency interval.

\begin{figure}
\includegraphics[angle=0,width=0.48\textwidth]{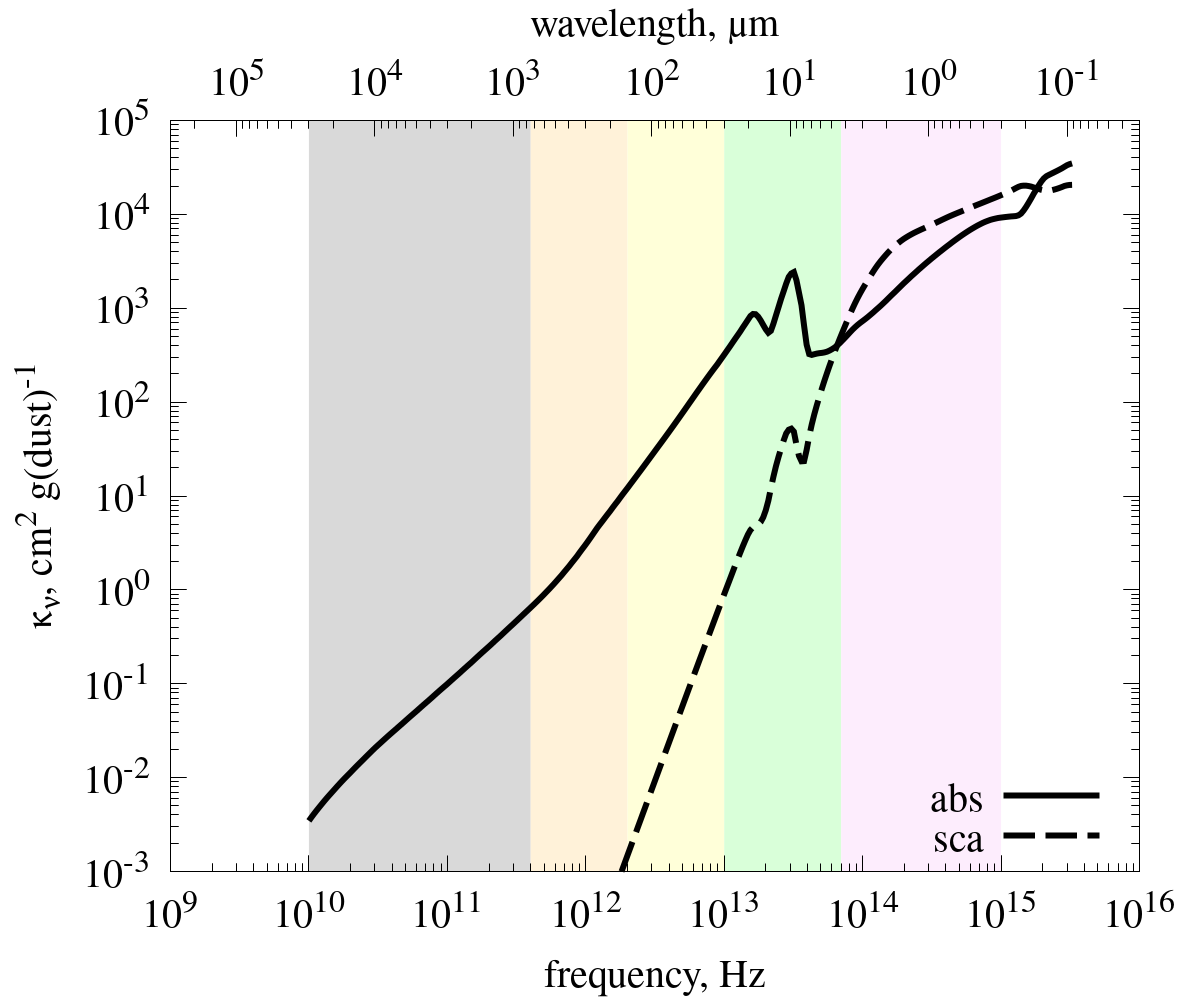}\\
\includegraphics[angle=0,width=0.23\textwidth]{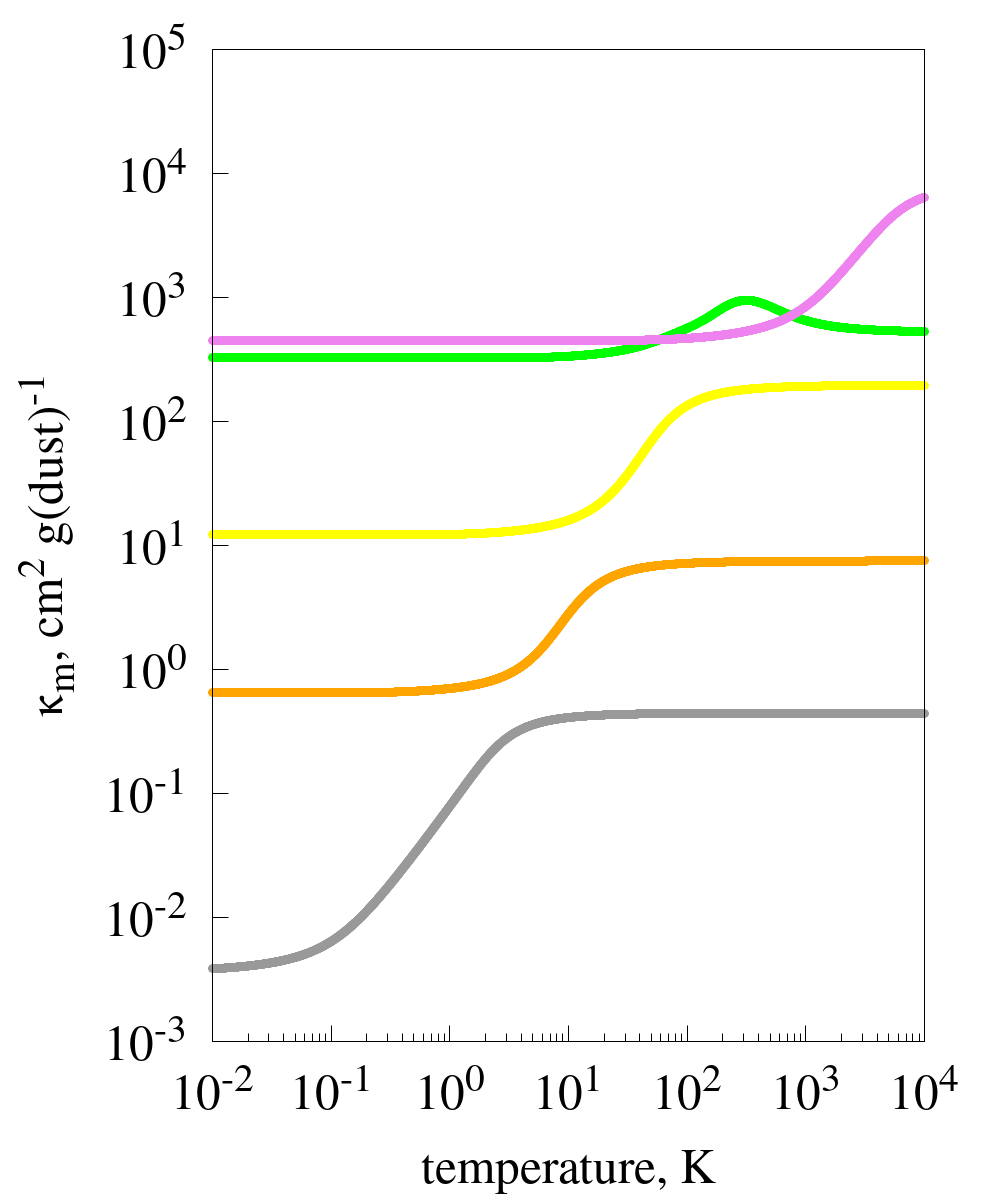}
\includegraphics[angle=0,width=0.23\textwidth]{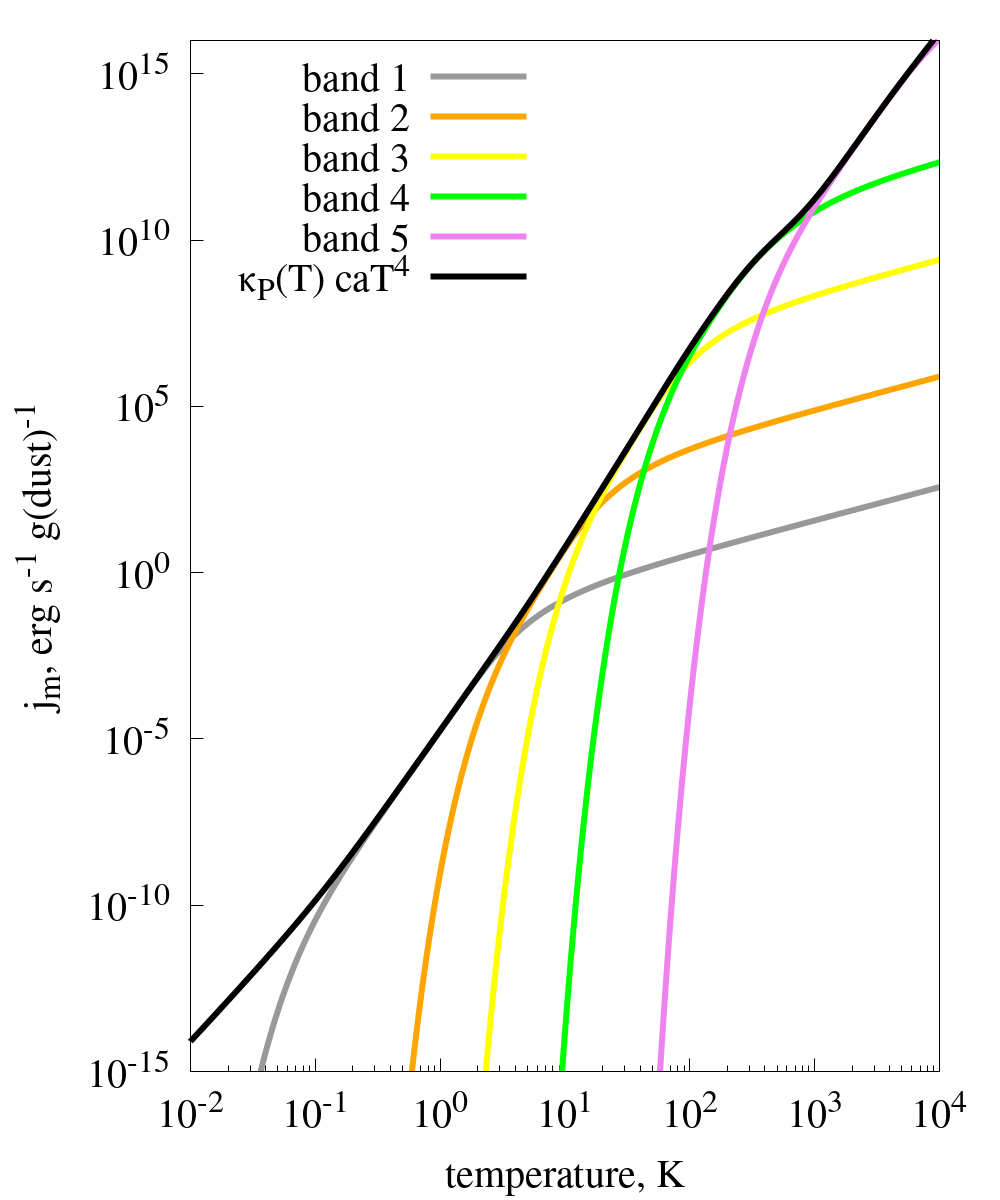}
\caption{Upper panel: absorption and scattering coefficients as functions
of frequency. The spectral channels are highlighted in color. Lower row:
dependencies of the absorption coefficients $\kappa_m$ and emission
coefficients $j_m$, integrated over frequency within the selected
frequency ranges, on temperature.}
\label{fig:opac}
\end{figure}

The interval-averaged coefficients $\alpha_m(T)$ and $\epsilon_m(T)$ are
calculated using the same spectral absorption coefficients
$\kappa_{\nu}^{\rm abs}$ as in Paper~I. The values of
$\kappa_{\nu}^{\rm abs}$ are obtained using Mie theory for a mixture of
spherical silicate and graphite dust grains (mass fraction of graphite
grains is 0.2), with a power-law size distribution $n(a)\propto a^{-3.5}$
and minimum and maximum grain radii of $5\times 10^{-7}$ and $10^{-4}$
cm.

The dependence of the absorption coefficient $\kappa_{\nu}^\text{abs}$ on
frequency is shown on the upper panel of Fig.~\ref{fig:opac}. We do not
take into account the scattering of thermal radiation, but note that in
the used dust model, absorption dominates over scattering up to
$\approx10^{14}$ Hz. This frequency corresponds to the maximum radiation
at a temperature of $\approx1700$ K, which is higher than the typical
values in protoplanetary disks in the bulk of the disk.

In the basic implementation of the mFLD$^{\rm s}$ method, the spectral
range is divided by us into five intervals, which are highlighted in
color on the upper panel of Fig.~\ref{fig:opac}. The dependencies of the
absorption coefficients $\kappa_m(T)=\alpha_m(T)/\rho_{\rm dust}$ and
emission coefficients $j_m(T)=c\epsilon_m(T)/\rho_{\rm dust}$ on
temperature, calculated per gram of dust, are shown on the lower panels
of Fig.~\ref{fig:opac}.

The differential operator describing the diffusion of radiation within
each frequency interval has the form:
\begin{equation}
\hat{\Lambda}E_m = -{\nabla}\,{\bf F}_m
= {\nabla} \left( \frac{c\lambda_m}{\alpha_m}{\nabla}\, E_m \right),
\label{eq:operE}
\end{equation}
where ${\bf F}_m$ is the flux of thermal radiation, $\alpha_m$ is the
interval-averaged absorption coefficient, and $\lambda_m$ is the flux
limiter. The calculation of $\lambda_m$ is carried out in accordance with
the formulas of FLD theory:
\begin{eqnarray}
&&{\bf R}_m = - \dfrac{\nabla E_m}{\alpha_m E_m} \label{eq:fld2} \\
&&\lambda_m=\dfrac{1}{R_m}\left(\coth R_m -\frac{1}{R_m} \right). \label{eq:fld3}
\end{eqnarray}

{
The system of equations~\eqref{therm_sys1b}--\eqref{therm_sys2b}
must be supplemented with boundary conditions. The choice of boundary conditions
depends on the specific problem. For spherically symmetric problems with
no radiation absorption in the inner cavity, the thermal radiation
exiting through the inner boundary of the computational domain
is fully compensated by incoming radiation from opposite walls.
Therefore, in this case we can use the condition:
\begin{equation}
\left. \frac{\partial E_m}{\partial R} \right|_{\text{in}} = 0.
\label{eq_bc1}
\end{equation}

When modeling circumstellar disks, thermal radiation from the inner
boundary of the disk may partially escape into the polar regions.
The amount of radiation returning to the disk from opposite walls will
depend on the height, position of the inner boundary, and other disk parameters,
and in general is difficult to predict. At the inner boundary
of circumstellar disks, we use the condition that the radiation flux
is proportional to the product of energy density and the speed of light:
\begin{equation}
\left. F_m\right|_{\text{in}} = -pc \left(E_m-E_m^\text{cmb}\right),
\label{eq_bc2}
\end{equation}
where $F_m$ is the radial component of the flux determined by
equation~\eqref{eq:operE}, and $E_m^{\text{cmb}}$ is the energy density
of the cosmic microwave background in the frequency interval $m$. The
coefficient $p$ is introduced phenomenologically and describes the
fraction of freely escaping radiation. For $p=0$, this condition reduces
to condition~\eqref{eq_bc1}. The value $p=1/2$ corresponds to the case
where radiation isotropically escapes the medium. The coefficient $p=1$
(used in our calculations) corresponds to the limiting case where
radiation freely escapes the medium perpendicular to the boundary.

For the outer boundary of the computational domain, we assume that thermal
radiation freely escapes the medium. In this case, the boundary condition
we use has the form:
\begin{equation}
\left. F_m\right|_{\text{out}} = c \left(E_m-E_m^\text{cmb}\right).
\label{eq_bc3}
\end{equation}
}

In the mFLD$^{\rm s}$ method, the heating of the disk by direct stellar
radiation is calculated using a ray-tracing method with spectral
absorption coefficients, similar to how it is implemented in the
FLD$^{\rm s}$ method (see Paper~I).

\subsection{Characteristic Local Times of the Radiative Transfer Equations}
Before proceeding to the description of the numerical method for solving
the equations implementing the FLD approximation, let us estimate the
characteristic times in the system \eqref{therm_sys1}--\eqref{therm_sys2}
(taking it as a simpler example). To do this, we will extract two
subsystems of equations from it. The first subsystem of equations:
\begin{eqnarray}
\rho c_{\rm V} \frac{\partial T}{\partial t}&=& c \rho\kappa \left(E - a T^4\right)
\label{therm_sys1c}\\
\frac{\partial E}{\partial t}&=& - c \rho\kappa \left(E - a T^4\right)
\label{therm_sys2c}
\end{eqnarray}
describes the energy exchange between the medium and the radiation field.
The representation $aT^4\approx aT^{3}_0\cdot T$ transforms
\eqref{therm_sys1c}--\eqref{therm_sys2c} into a system of linear ordinary
differential equations. The eigenvalue of this system is $\lambda_{\rm
th} = -\left(\dfrac{c\kappa\, aT^{3}_0}{c_{\rm V}}+c\rho\kappa\right)$.
The characteristic relaxation time of the solution $\Delta t_{\rm th} =
-1/\lambda_{\rm th}$ can be represented as:
\begin{eqnarray}
\dfrac{1}{\Delta t_{\rm th}} = \dfrac{1}{\Delta t_{\rm th,1}} + \dfrac{1}{\Delta t_{\rm th,2}},
\end{eqnarray}
where:
\begin{eqnarray}
&&\Delta t_{\rm th,1} = \dfrac{c_{\rm V}}{c\kappa\,aT^{3}_0}, \\
&&\Delta t_{\rm th,2} = \dfrac{1}{c\rho\kappa}.
\end{eqnarray}
The times $\Delta t_{\rm th,1}$ and $\Delta t_{\rm th,1}$ can be
considered as characteristic times of energy transfer between the matter
and the radiation field. In this notation, it is clear that the total
time $\Delta t_{\rm th}$ is less than either of the two $\Delta t_{\rm
th,1}$ and $\Delta t_{\rm th,2}$ (by analogy with the total resistance of
parallel-connected resistors).

The second subsystem of equations is written by taking the operator
$\hat{\Lambda} E$ as one-dimensional and using the Eddington
approximation $\lambda=1/3$:
\begin{equation}
\frac{\partial E}{\partial t} =
\dfrac{\partial}{\partial x}\left(\dfrac{c}{3\kappa\rho}\dfrac{\partial E}{\partial x} \right).
\label{eq:dif}
\end{equation}
This is the classical diffusion equation, in this case describing the
redistribution of radiation energy in space. The characteristic diffusion
time $\Delta t_{\rm d}$ for \eqref{eq:dif} depends on the spatial scale
$h$:
\begin{equation}
\Delta t_{\rm d} = \dfrac{3\rho\kappa}{c} h^2.
\end{equation}
Note that $\Delta t_{\rm d}$ is also the maximum time step that ensures
the stability of the numerical solution (if $h$ is the minimum cell size)
when using an explicit scheme for integrating~\eqref{eq:dif}.

When using explicit methods for integrating the
system~\eqref{therm_sys1}--\eqref{therm_sys2}, the maximum time step must
be less than the characteristic times $\Delta t_{\rm th,1}$, $\Delta
t_{\rm th,2}$, and $\Delta t_{\rm d}$ obtained above. Otherwise, the
numerical solution may be either unstable or give unphysical (negative)
values. Let us estimate the characteristic times, taking the conditions
deep inside the protoplanetary disk: $\rho=10^{-13}$ g cm$^{-3}$,
$T_0=100$ K, $\kappa=1$ cm$^2$ g$^{-1}$ (gas), and the minimum grid step
$h=0.01$ au:
\begin{eqnarray}
&&\Delta t_{\rm th,1} \approx 10^{-2}\text{ years,}\\
&&\Delta t_{\rm th,2} \approx 10^{-5}\text{ years,}\\
&&\Delta t_{\rm d} \approx  10^{-8}\text{ years.}
\end{eqnarray}
The minimum of these values is many orders of magnitude smaller than the
characteristic dynamic times ($\sim 10^{-1}$ years) for the inner parts
of the disk (coinciding with the limitation on the maximum time step when
solving hydrodynamic equations), which makes the use of direct explicit
schemes for approximating the system
\eqref{therm_sys1}--\eqref{therm_sys2} in hydrodynamic problems extremely
inefficient. The same conclusions can be drawn for the system
\eqref{therm_sys1b}--\eqref{therm_sys2b}. One solution to this problem is
to use implicit schemes for approximating the equations of the thermal
model. Our implementation of this approach is described in the next
section.

\subsection{Numerical Solution Method}

Equations~\eqref{therm_sys1b}--\eqref{therm_sys2b} form a nonlinear
system of ($M$+1) (where $M$ is the number of spectral intervals) partial
differential equations of the diffusion type. Its solution is found in an
axisymmetric approximation using a spherical coordinate system ($R,
\theta$). The spatial grid, approximation of differential operators, and
the principle of constructing the method are completely analogous to
those described in detail in Paper~I, except for the need to
simultaneously include $M$ equations~\eqref{therm_sys2b}, describing the
diffusion of $M$ components of radiation energy, instead of
one~\eqref{therm_sys2}. Therefore, we will only briefly describe the
modification of the method.

The solution uses an implicit finite-difference method, in which the
exchange terms $\alpha_m(T)E_m$ and $\epsilon_m(T)$ on the right-hand
side of equations~\eqref{therm_sys1b}--\eqref{therm_sys2b}, as well as
the differential operator \eqref{eq:operE}, depend on the values of the
functions at the new ($n+1$) time layer:
\begin{eqnarray}
\rho c_{\rm V} \frac{T-T^{n}}{\Delta t}&=& c \sum_{m=1}^{M}\left[\alpha_m(T) E_m - \epsilon_m(T)\right] + S,
\label{num_sys1b} \\
\frac{E_m-E_m^{n}}{\Delta t}&=& - c\left[\alpha_m(T) E_m - \epsilon_m(T)\right] + \hat{\Lambda}_m E_m,
\label{num_sys2b}
\end{eqnarray}
where $T^{n}$ and $E_m^{n}$ are the values from the $n$-th time layer,
$T$ and $E_m$ are the sought values at the time layer ($n+1$) for a given
spatial cell. In the equations above, the lower spatial indices are
omitted for brevity --- for all quantities, they correspond to the
considered cell $(i,j)$, except for the operator $\hat{\Lambda}_m$, which
connects the cell $(i,j)$ with four adjacent cells in radius $R$ and
angle~$\theta$.

To solve the system of equations of the thermal evolution of the
medium~\eqref{num_sys1b}--\eqref{num_sys2b}, an iterative process over
$k$ is organized, which is a combination of the Newton method and the
simple iteration method:
\begin{eqnarray}
\rho c_{\rm V} \frac{T^{k+1}-T^{n}}{\Delta t}= c \sum_{m=1}^{M}\left[\alpha_m(T^{k}) E^{k+1}_m - \epsilon_m(T^{k+1})\right] + S,
\label{num_sys1c} \\
\frac{E_m^{k+1}-E_m^{n}}{\Delta t}= - c\left[\alpha_m(T^k) E^{k+1}_m - \epsilon_m(T^{k+1})\right] + \hat{\Lambda}_m E^{k+1}_m,
\label{num_sys2c}
\end{eqnarray}
where $T^{k}$ is the temperature value from the previous iteration,
$T^{k+1}$, $E^{k+1}_m$ are the sought values. In this case, the emission
coefficients entering the right-hand sides of the equations are
linearized using the approximation:
\begin{equation}
\epsilon_m(T^{k+1})\approx \epsilon_m(T^k)
+\left(\dfrac{\partial \epsilon_m}{\partial T}\right)_{T^k} \left(T^{k+1}-T^k\right).
\label{eq:linear}
\end{equation}
After substituting the expressions for the spatial operators
$\hat{\Lambda}_m E^{k+1}_m$, connecting the current cell with four
neighboring ones, and transforming the terms, the system of linear
algebraic equations~\eqref{num_sys1c}--\eqref{eq:linear}, {
supplemented by finite-difference approximations of the boundary
conditions \eqref{eq_bc2}--\eqref{eq_bc3},} can be written as:
\begin{equation}
\bf\hat{H}Y=G,
\label{eq:hyper}
\end{equation}
where $\bf Y$ is the hypervector of unknown variables:
\begin{equation}
{\bf Y} = \left({\bf y}(1,1),\,\, {\bf y}(1,2),\,\, ...,\,\, {\bf y}(N_{R},N_{\theta})\right)^{T},
\end{equation}
each component of which is a vector and contains the unknown quantities
for the corresponding cell, i.e.
\begin{equation}
{\bf y}(i,j) = \left(T^{k+1}(i,j),\,\, E_1^{k+1}(i,j),\,\, ...,\,\, E_M^{k+1}(i,j)\right)^T.
\end{equation}
\begin{figure}
\centering
\includegraphics[angle=0,width=0.48\textwidth]{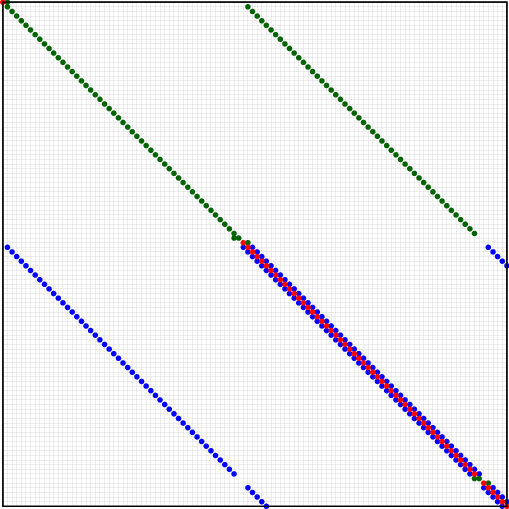}\\
\includegraphics[angle=0,width=0.23\textwidth]{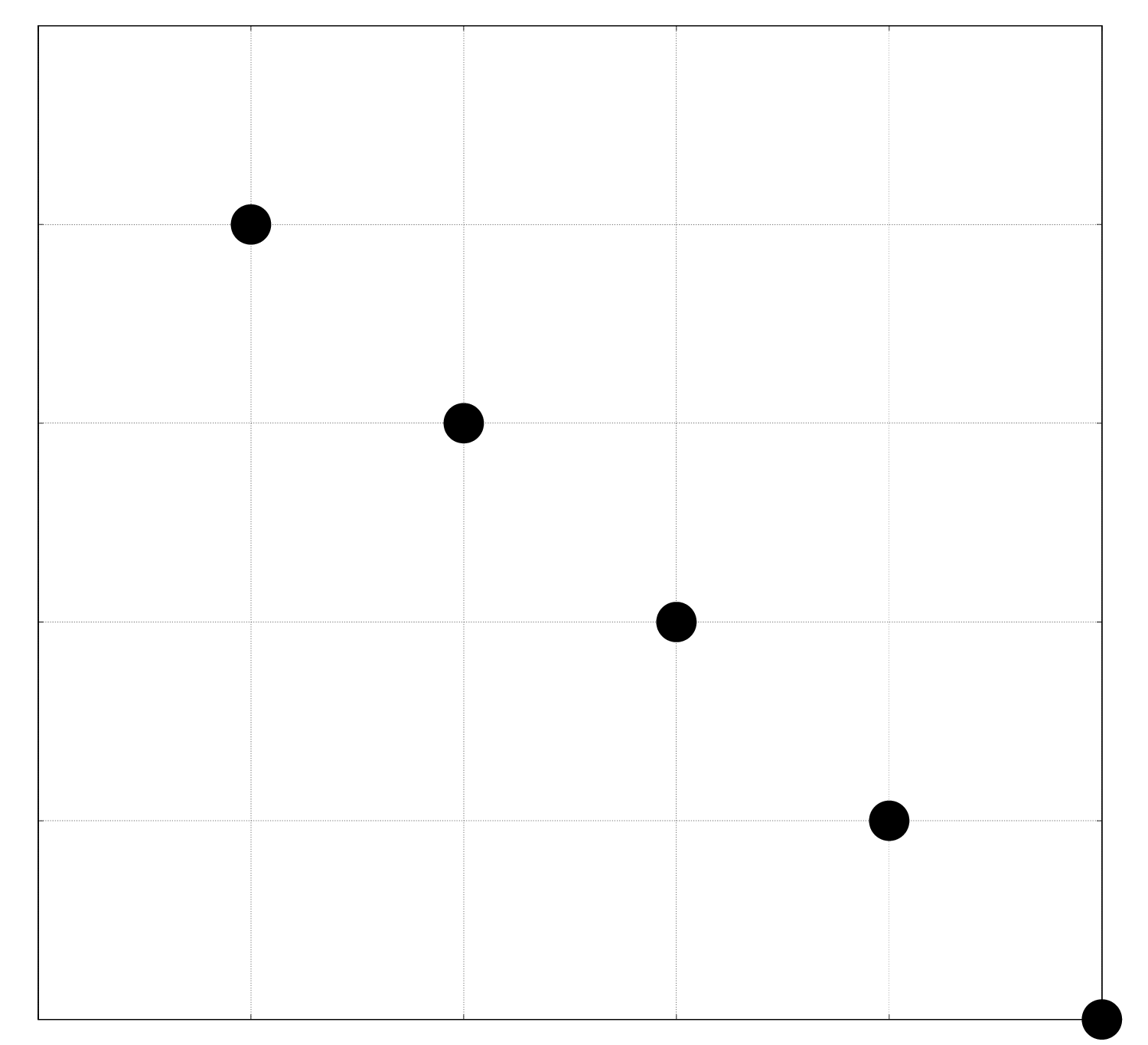}
\includegraphics[angle=0,width=0.23\textwidth]{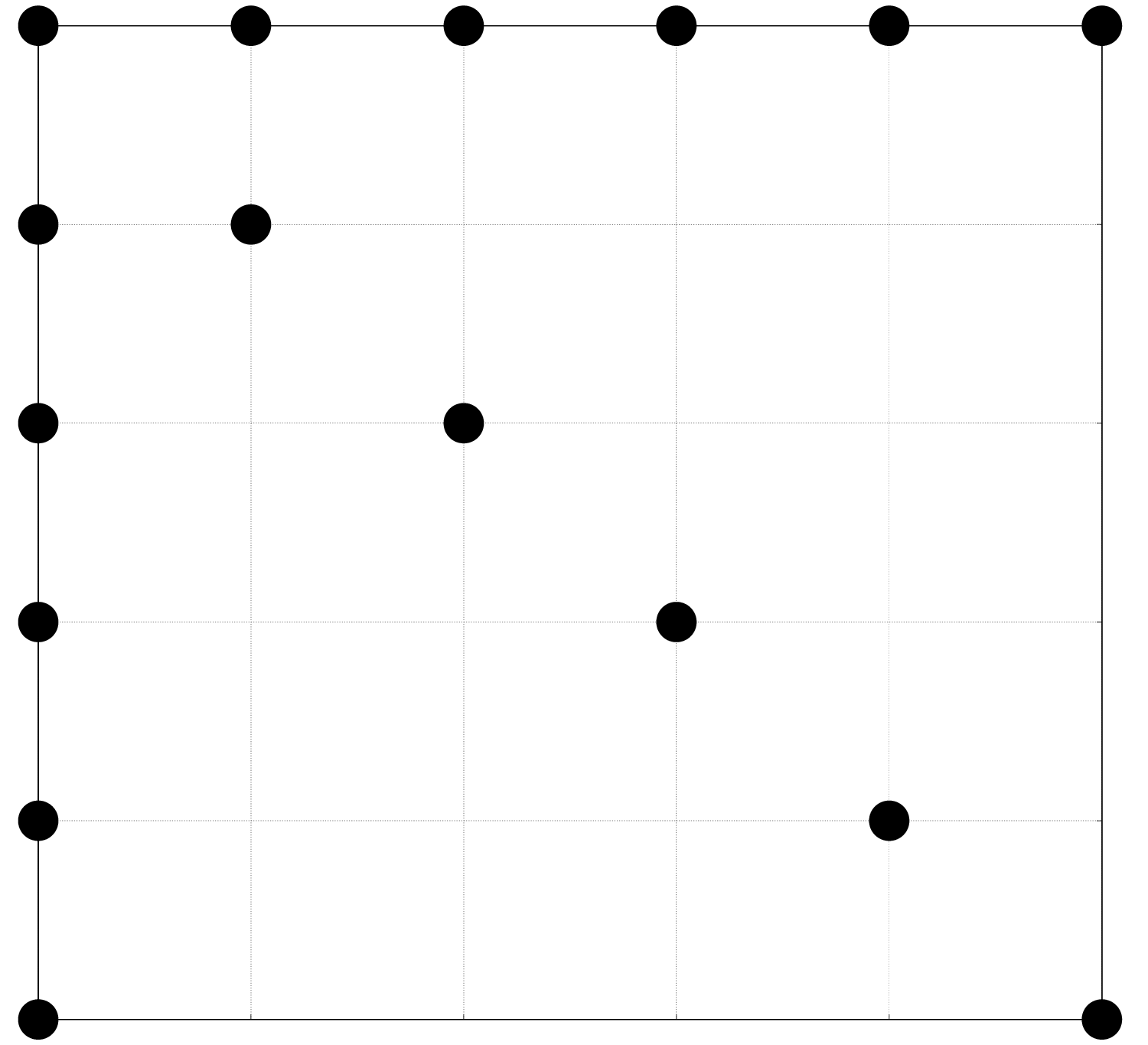}
\caption{Top: structure of the hypermatrix, showing the first
$110\times110$ of $5200\times5200$ elements (the total number of rows of
the hypermatrix is $N_R(N_{\theta} + 2) = 100(50+2) =5200$, where $N_R=100$
and $N_{\theta}=50$ are the number of grid cells in radius and angle,
respectively). Each colored dot corresponds to its non-zero matrix. Blue dots
denote matrices $\hat{A}$, $\hat{B}$, $\hat{D}$, $\hat{E}$ connecting the current cell with neighboring
ones. Red dots show matrices $\hat{C}$, corresponding to the current cell. Green dots
show matrices implementing boundary conditions. Bottom row: structure of
matrices $\hat{A}$, $\hat{B}$, $\hat{D}$, $\hat{E}$ (left panel) and $\hat{C}$ (right panel), which are elements
of the hypermatrix.}
\label{fig:matrix}
\end{figure}
In the above expressions, $N_{R}$ and $N_{\theta}$ are the number of
cells in radius and angle, respectively, and the symbol $T$ above the
brackets is the transposition sign. The hypermatrix of the system $\bf
\hat{H}$ has the form schematically shown in Fig.~\ref{fig:matrix}. The
colored dots in this diagram represent non-zero elements. The blue
dots show the matrices (let's call them $\hat{A}$, $\hat{B}$,
$\hat{D}$, $\hat{E}$), connecting the current cell with four neighboring
ones (in radius and angle). The red dots show the matrices for the
current cell (let's call them $\hat{C}$) -- they connect the local
unknowns $T$ and $E_1$, ..., $E_M$. The structure of the matrices
$\hat{A}$, $\hat{B}$, $\hat{D}$, $\hat{E}$, and $\hat{C}$ is shown on the
lower panels of Fig.~\ref{fig:matrix}, where dots mark non-zero
values. Note that the use of the hypermatrix form of representing the
system of linear equations is extremely convenient for further solution
and has been used previously, for example, in the Feautrier method for
solving the radiation transfer equation in stellar atmospheres,
see~\cite{1978stat.book.....M}.

To solve the system of equations \eqref{eq:hyper}, written in hypermatrix
form, we use a modification of the Gauss method presented in Paper~I,
transforming it into a hypermatrix form. The solution method (algorithm
for transforming rows and reducing the hypermatrix of the system to an
upper triangular form) is similar to that described in Paper~I, except
that the operations are performed with matrices (elements of the
hypermatrix of the system), and not with numbers (elements of the system
matrix).

The iterative process over $k$, within which the system of linear
equations is solved, is carried out until convergence --- usually the
iterations converge in a few steps. After that, the values at the new
time layer are declared found: $T^{n+1} = T^{k+1}$, $E^{n+1}_m =
E^{k+1}_m$.

For the calculations presented in this article, we use a non-uniform
grid, condensing in $R$ towards the center and in $\theta$ towards the
midplane, with a resolution of 100 radial by 50 angular cells.
The grid structure is shown on the left upper panel of
Fig.~\ref{fig:compareA3}.

\section{Testing and Analyzing the Two-Dimensional Method}

\subsection{Testing the Two-Dimensional Method in Spherical Symmetry Mode}

Consider the problem of heating a homogeneous spherically symmetric shell
by a central star. The parameters of the shell: inner radius 0.5 au,
outer radius 250 au, molecular hydrogen concentration $10^9$~cm$^{-3}$.
The star is modeled as a blackbody with an effective temperature of 3800 K
and a radius of 1.9 $R_{\odot}$. Note that a homogeneous shell does
not correspond to any astrophysical object; we chose it solely to exclude
the influence of inhomogeneity on possible grid effects. We will be
interested in the stationary temperature distribution, which, with these
parameters, is reached in $\approx100$ years. As a reference solution,
the results of calculations using the RADMC-3D
code\footnote{\url{https://www.ita.uni-heidelberg.de/~dullemond/software/radmc-3d/index.php}}~\citep{2012ascl.soft02015D}
will be used.

\begin{figure}
\includegraphics[angle=0,width=0.23\textwidth]{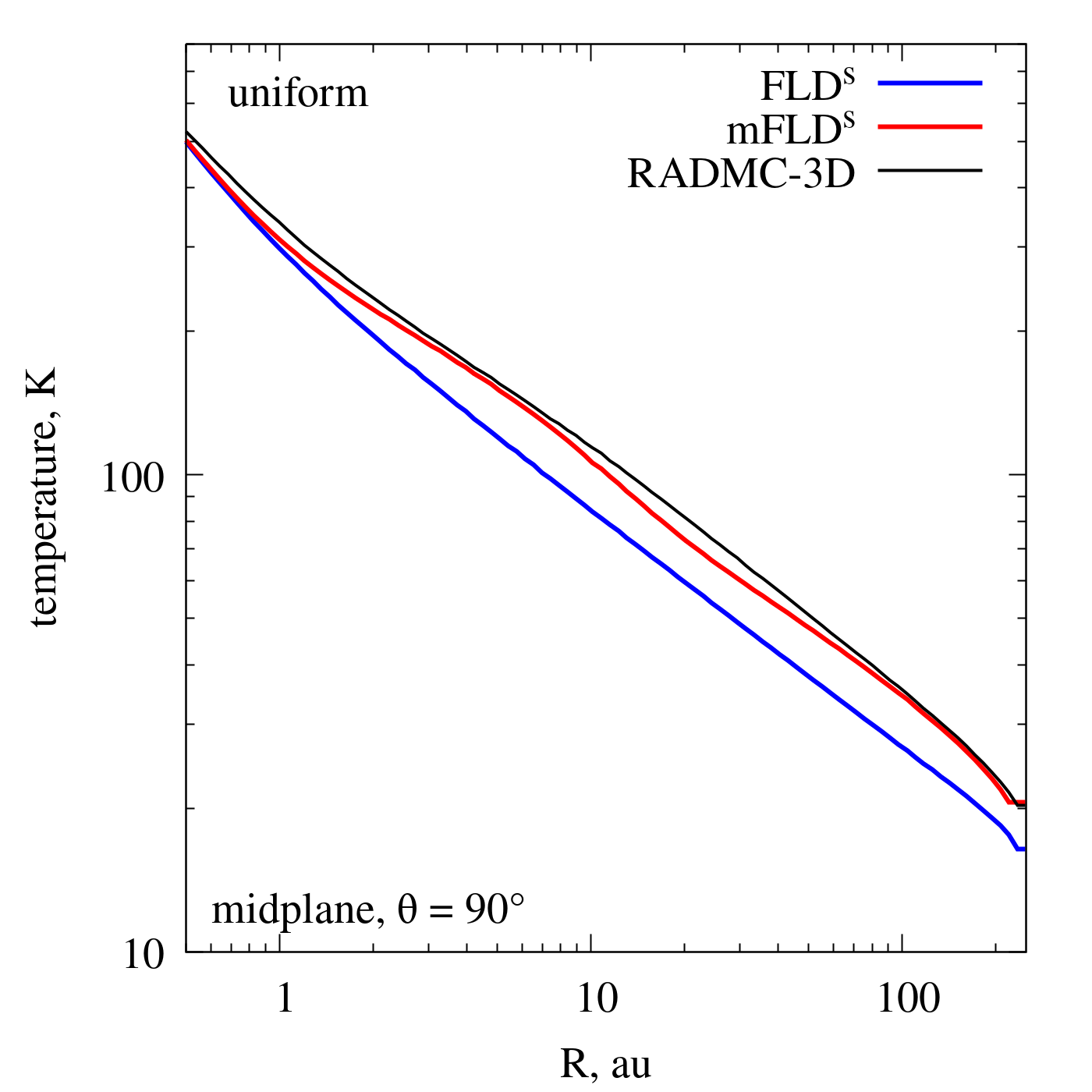}
\includegraphics[angle=0,width=0.23\textwidth]{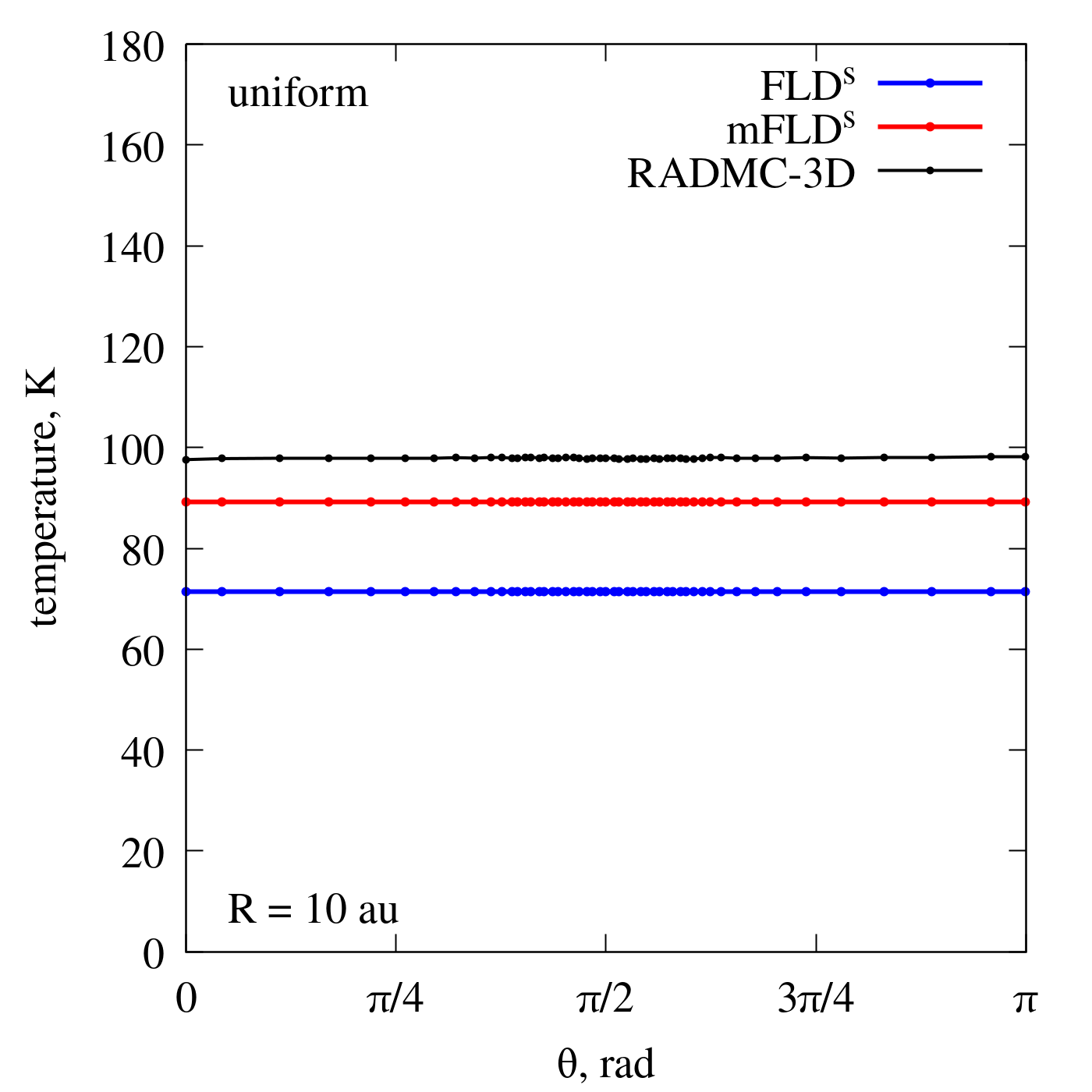}\\
\includegraphics[angle=0,width=0.23\textwidth]{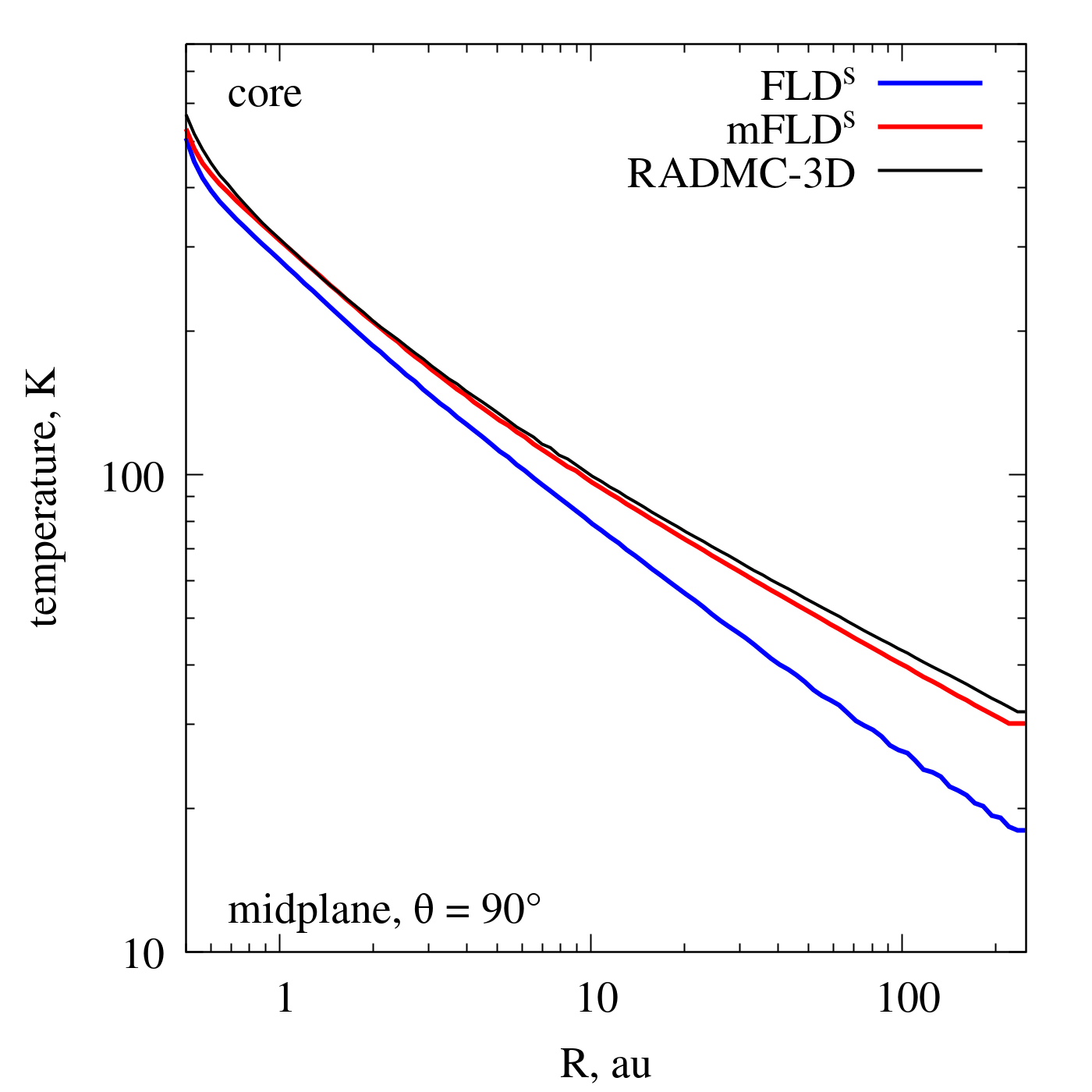}
\includegraphics[angle=0,width=0.23\textwidth]{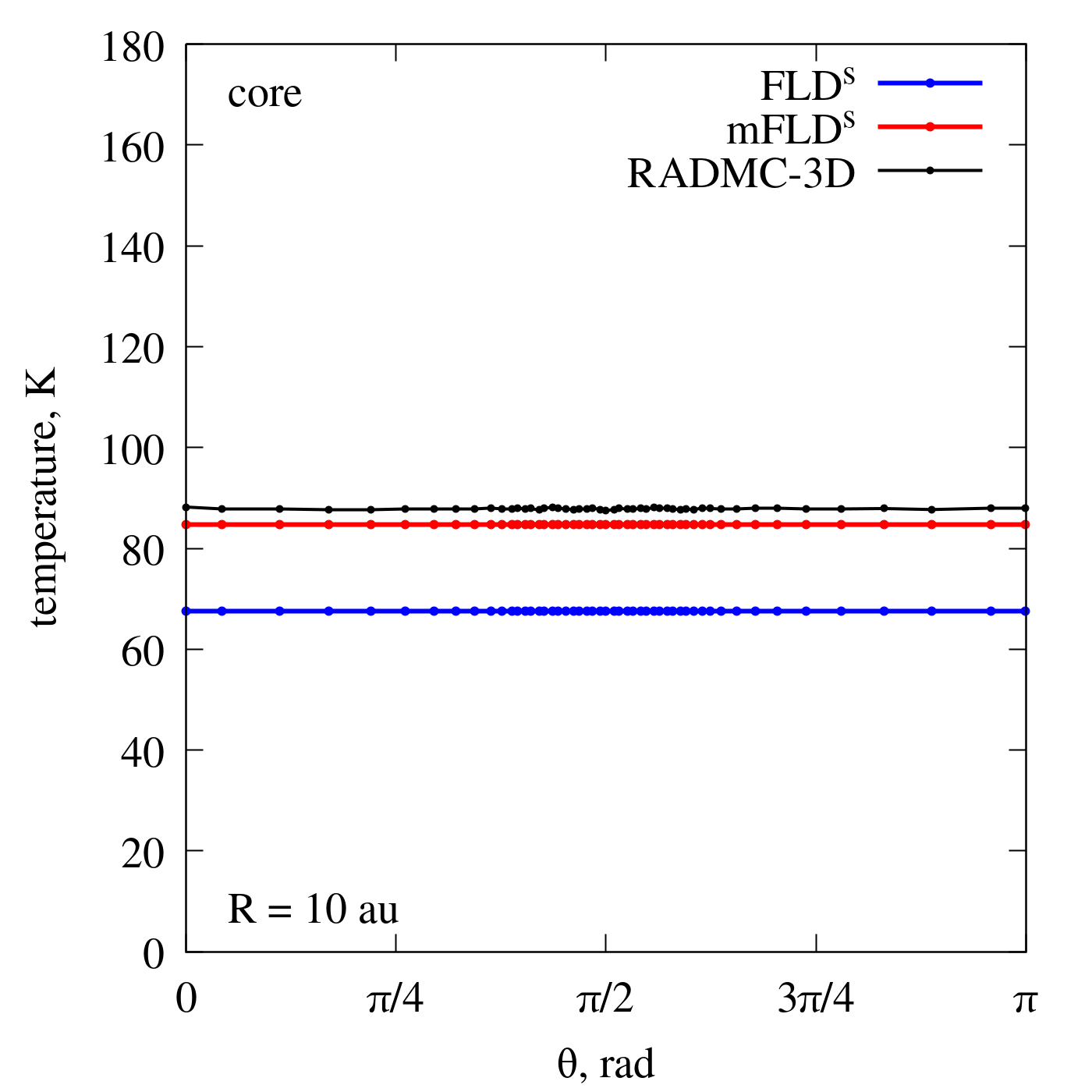}
\caption{Stationary temperature distributions for a model of a
homogeneous (upper row) and inhomogeneous (lower row) spherically
symmetric cloud, obtained by the FLD$^{\rm s}$, mFLD$^{\rm s}$ methods,
and the RADMC-3D code.}
\label{fig:compareA1}
\end{figure}

Figure~\ref{fig:compareA1} shows the stationary temperature distributions
obtained by the FLD$^{\rm s}$, mFLD$^{\rm s}$ methods, and the RADMC-3D
code. It can be seen that the temperature distributions obtained by
mFLD$^{\rm s}$ agree significantly better with the results of RADMC-3D
than the distributions of FLD$^{\rm s}$. All three considered methods
well preserve spherical symmetry, as can be seen from the temperature
distribution along the angle $\theta$ (at a distance of 10 au from the
center) on the right panel of Fig.~\ref{fig:compareA1}.

The lower panels of Fig.~\ref{fig:compareA1} also show a comparison of
temperature distributions for a spherically symmetric inhomogeneous cloud
(a prototype of a protostellar shell) with a gas concentration
distribution $n(R)=10^{10}(R/0.5\,\mbox{au})^{-1.5}$ cm$^{-3}$. The
closeness of the results between mFLD$^{\rm s}$ and RADMC-3D {is
preserved for a wide range} of parameters of spherically symmetric shells
(optically thick and thin {to their own thermal radiation}, homogeneous
and inhomogeneous), while the FLD$^{\rm s}$ method gives significant (up
to 50\%) deviations from the reference distribution for shells that are
optically thick to stellar radiation but optically thin to their own
thermal radiation.

\subsection{Geometric Test: Homogeneous Cloud with a Heat Source Shifted Along the Polar Axis}

\begin{figure}
\includegraphics[angle=0,width=0.26\textwidth]{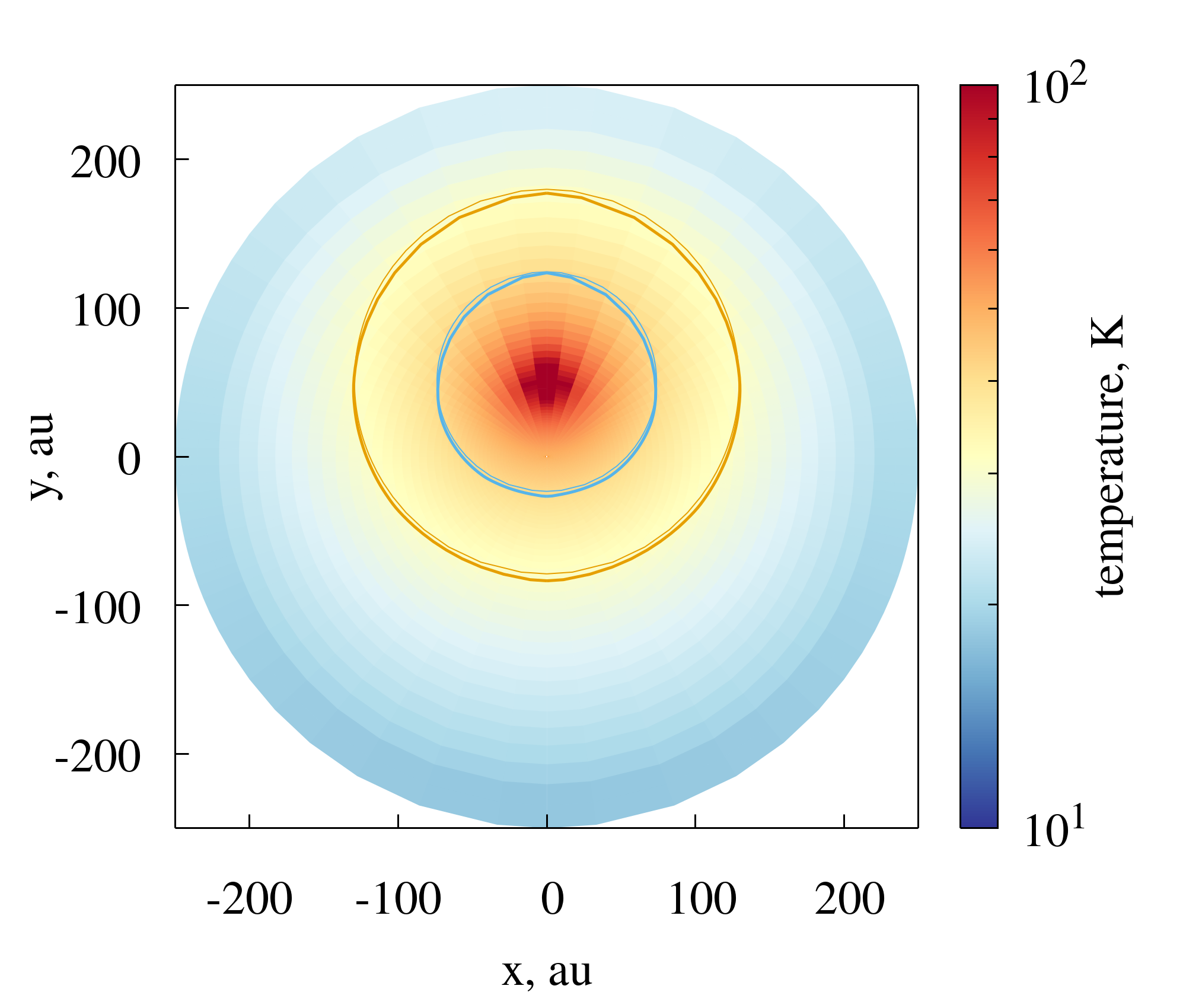}
\includegraphics[angle=0,width=0.21\textwidth]{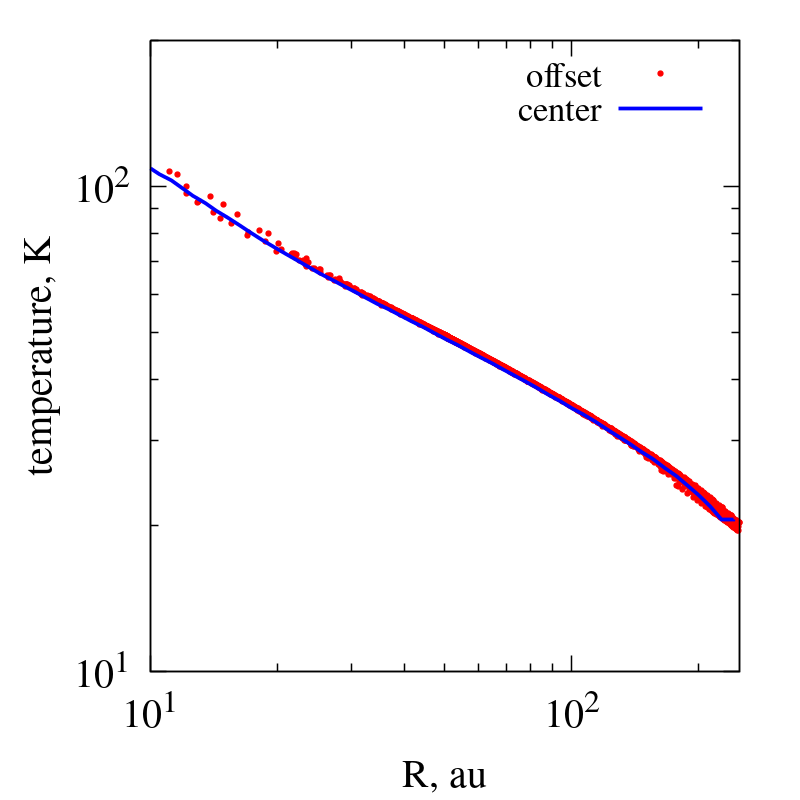}
\caption{Left panel: two-dimensional stationary temperature distribution
in the polar section of a homogeneous cloud with a heat source shifted
along the polar axis. Bold lines show the corresponding thermal map
levels of 30 and 40 K, thin lines show the same levels for the model with
the source at the origin, but shifted for ease of comparison. Right
panel: comparison of stationary temperature distributions for the shifted
and non-shifted sources. The results are obtained by the mFLD$^{\rm s}$
method.}
\label{fig:compareA2}
\end{figure}

Figure~\ref{fig:compareA2} shows the results of calculating the
stationary temperature in a homogeneous cloud with the parameters from
the previous section, where the heat source is shifted by 50 au from the
coordinate center along the polar axis. Heating is carried out by setting
the function $S$ in one cell adjacent to the polar axis. The heating
power corresponds to the luminosity of the star. The difference grid, as
in the previous calculation, is non-uniform both in radius and in angle.

The two-dimensional temperature distribution in the polar section of the
cloud (left panel of Fig.~\ref{fig:compareA2}) shows that the obtained
distribution is visually symmetric with respect to the heat source. A
detailed comparison between the temperature dependencies on the heating
center for the shifted and non-shifted sources (right panel of
Fig.~\ref{fig:compareA2}) allows us to claim a good agreement between the
distributions. The results of this test confirm the
correctness of the approximation of the original system of equations in a
curvilinear coordinate system on a non-uniform grid.

\subsection{Stationary Thermal Structure of a Protoplanetary Disk}
\begin{figure}
\includegraphics[angle=0,width=0.23\textwidth]{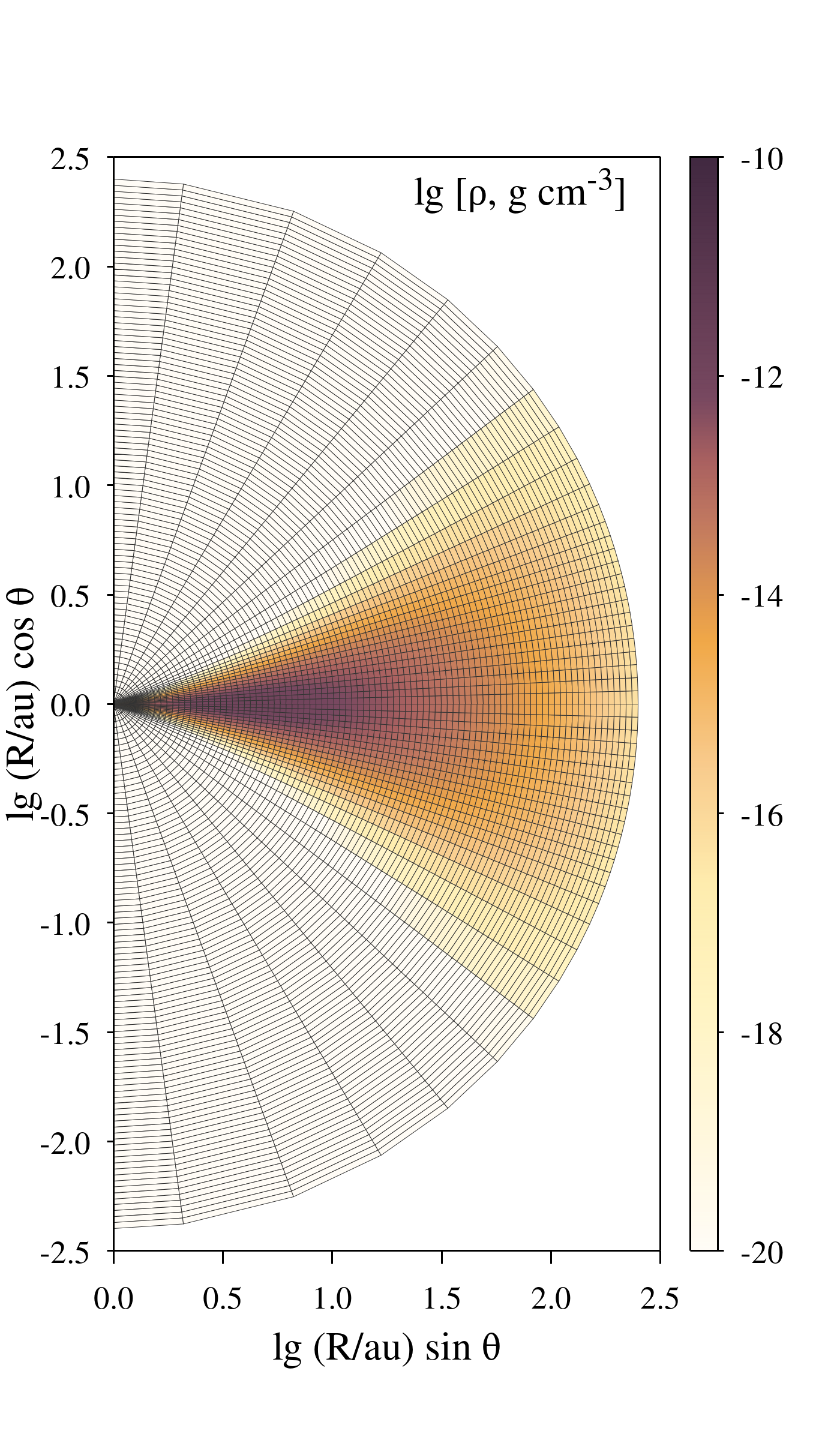}
\includegraphics[angle=0,width=0.23\textwidth]{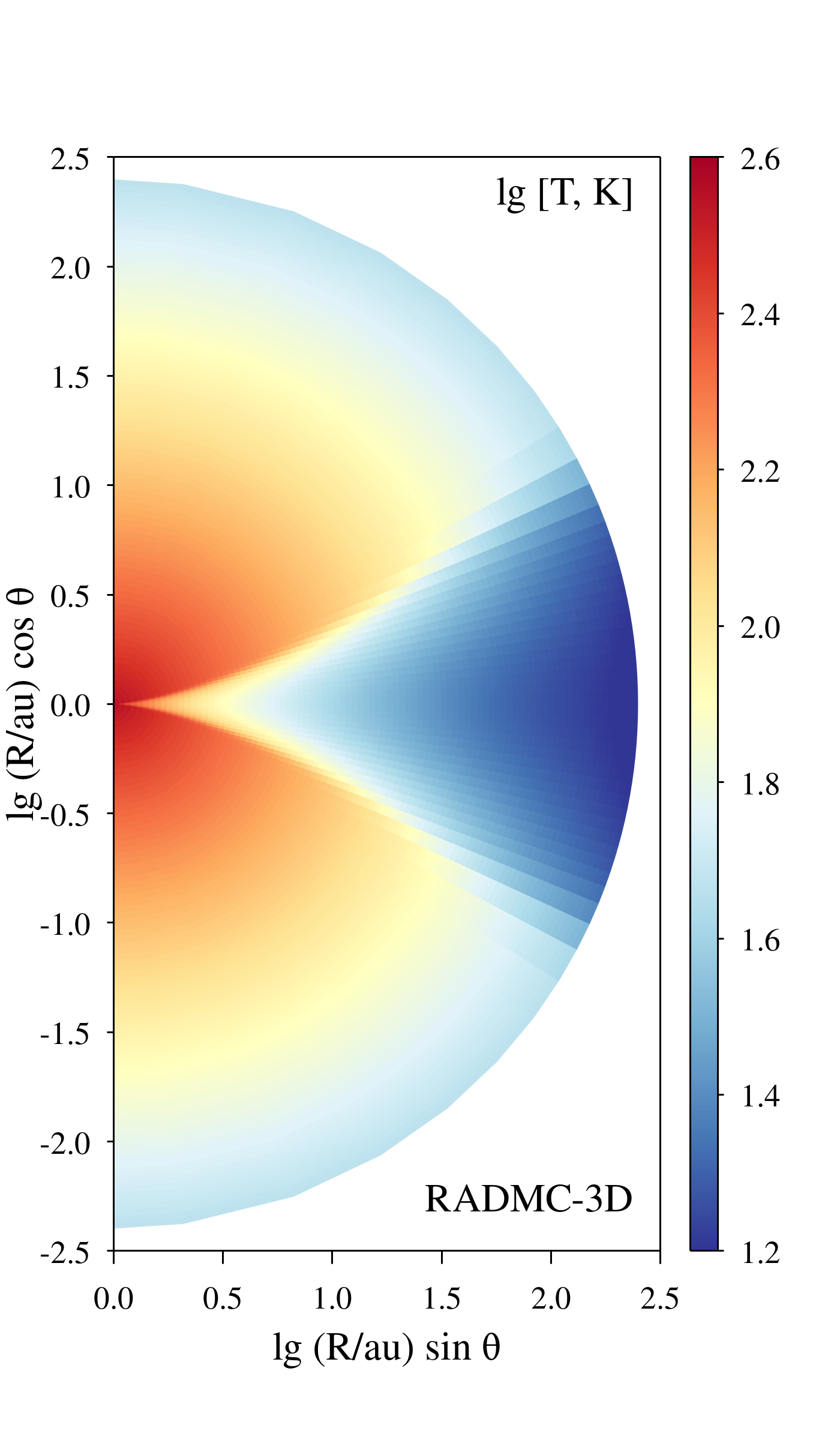}\\
\includegraphics[angle=0,width=0.23\textwidth]{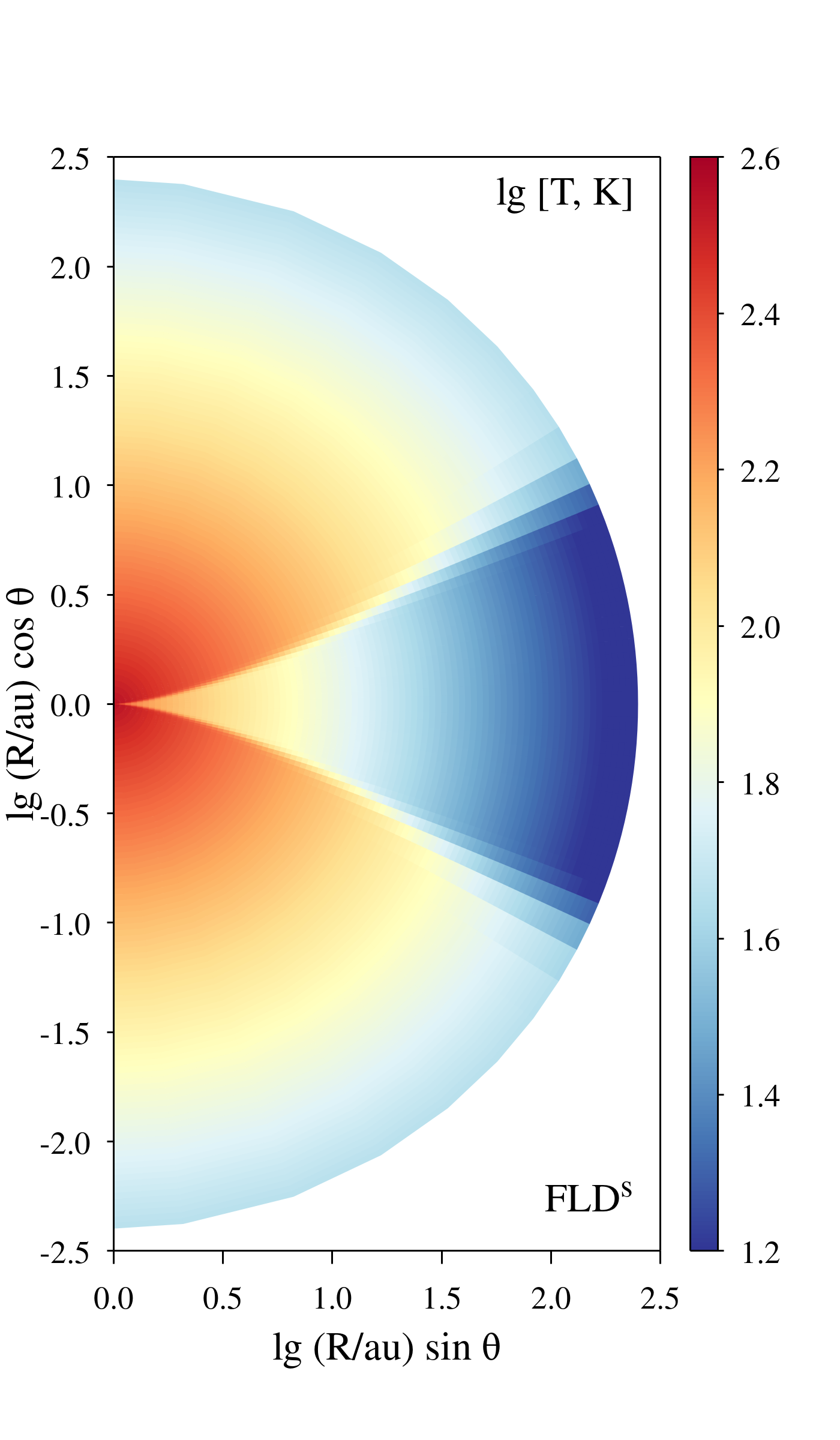}
\includegraphics[angle=0,width=0.23\textwidth]{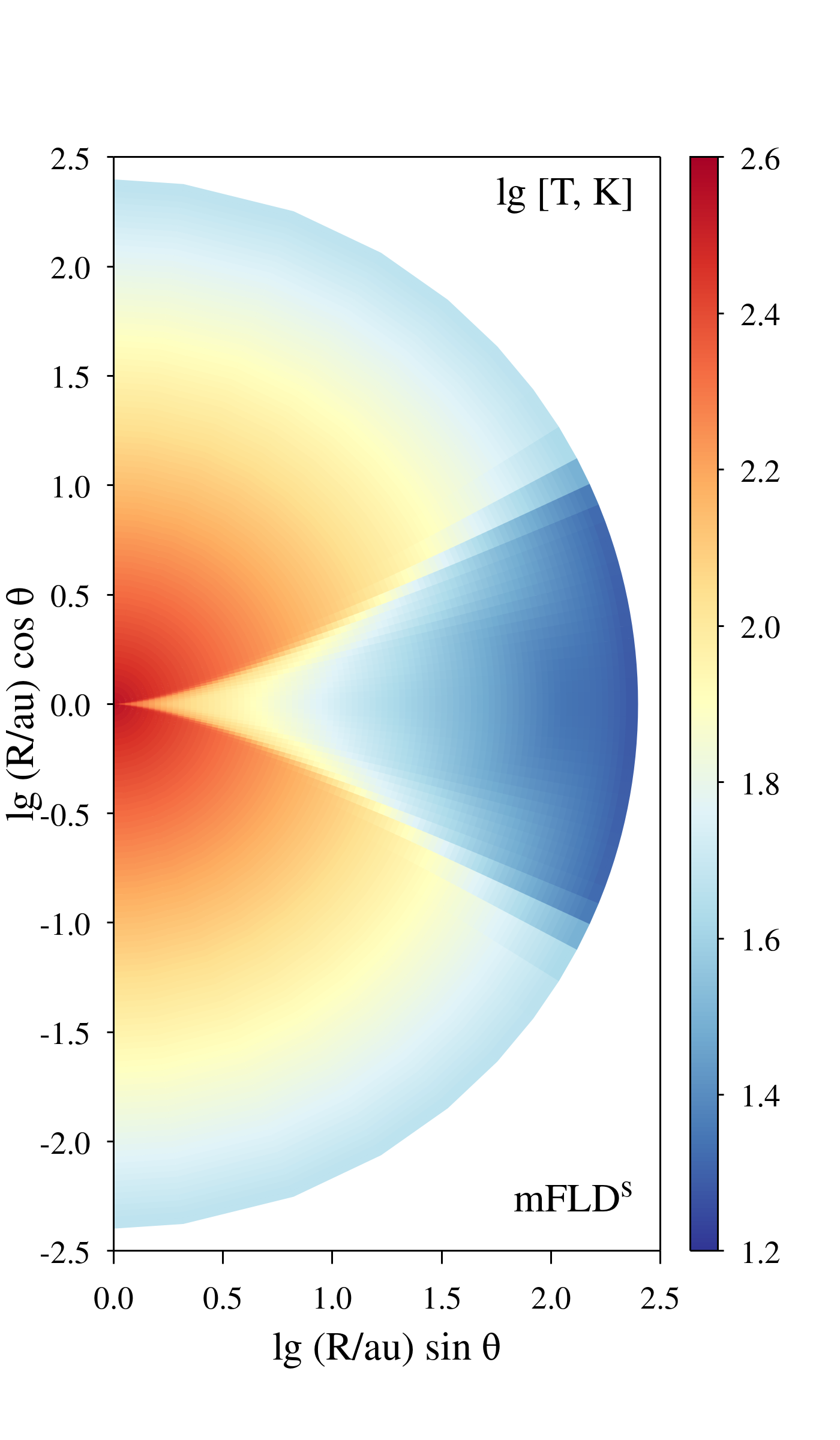}
\caption{Top left: density distribution for the gas-dust disk model (only
the right part from the polar axis is shown). Gray lines show the
boundaries of the grid cells. Top right: temperature distribution in the
polar section of the disk, obtained by the RADMC-3D code. Stationary
temperature distributions obtained by the FLD$^{\rm s}$ (bottom left) and
mFLD$^{\rm s}$ (bottom right) methods.}
\label{fig:compareA3}
\end{figure}
\begin{figure}
\vspace{0.6cm}
\includegraphics[angle=0,width=0.23\textwidth]{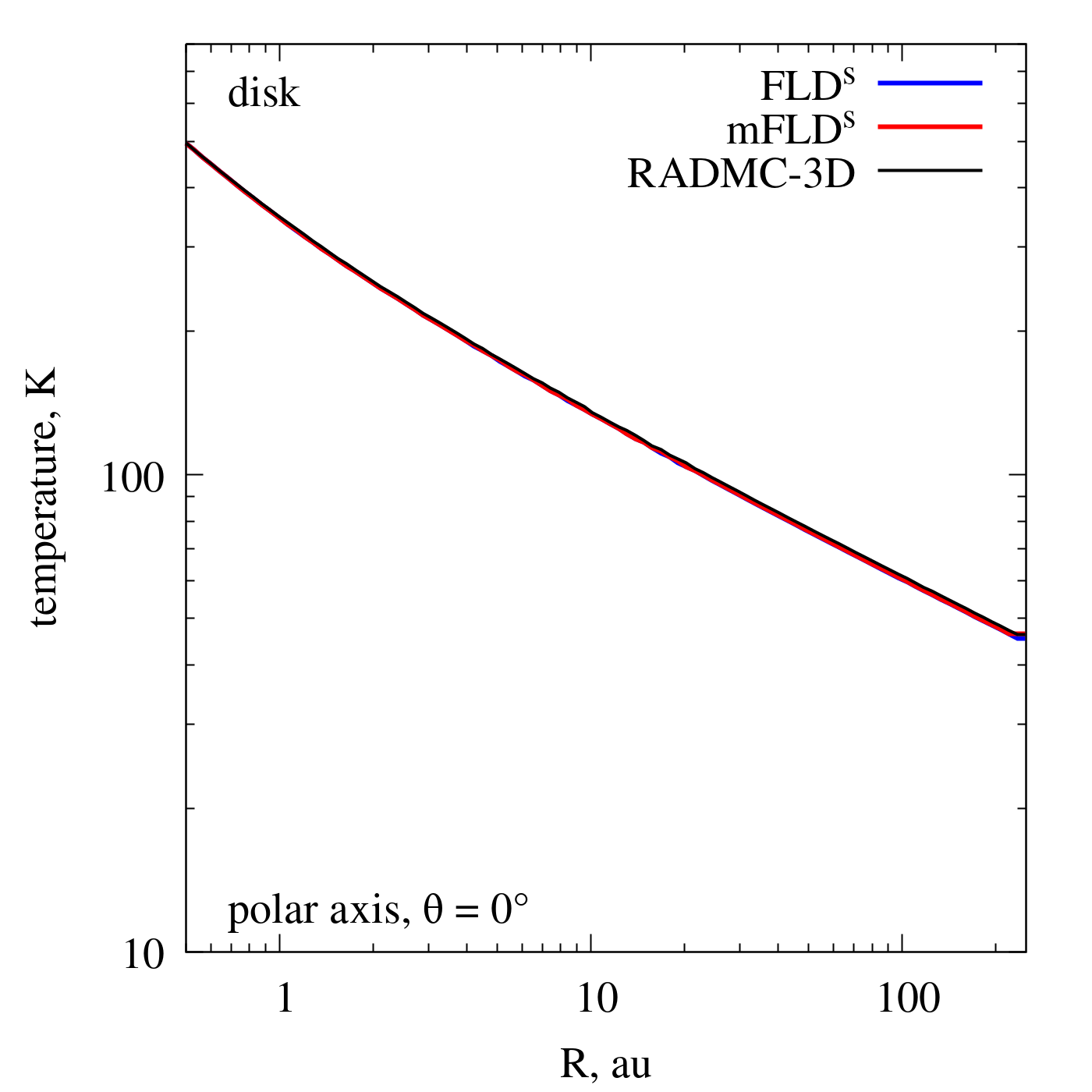}
\includegraphics[angle=0,width=0.23\textwidth]{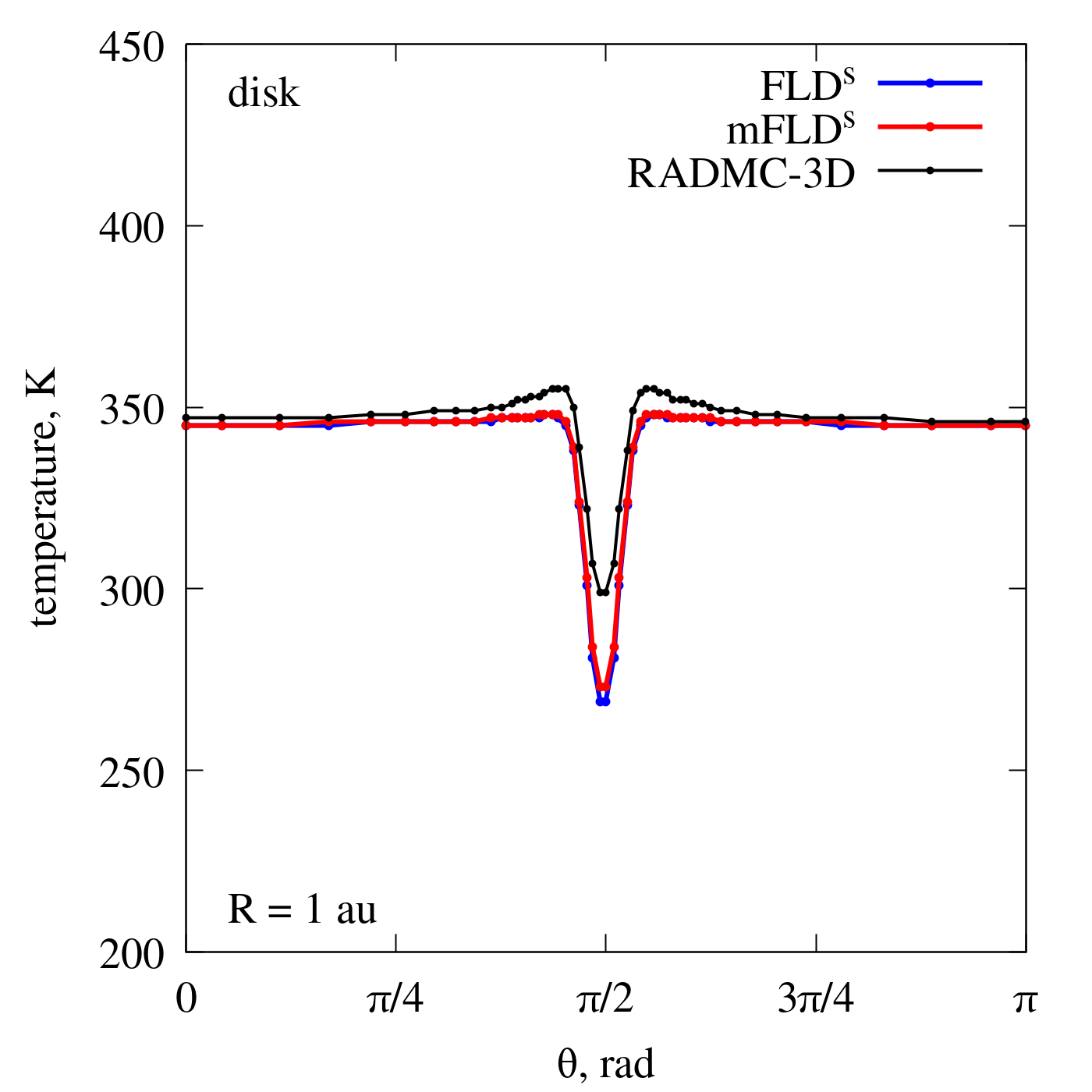}\\
\includegraphics[angle=0,width=0.23\textwidth]{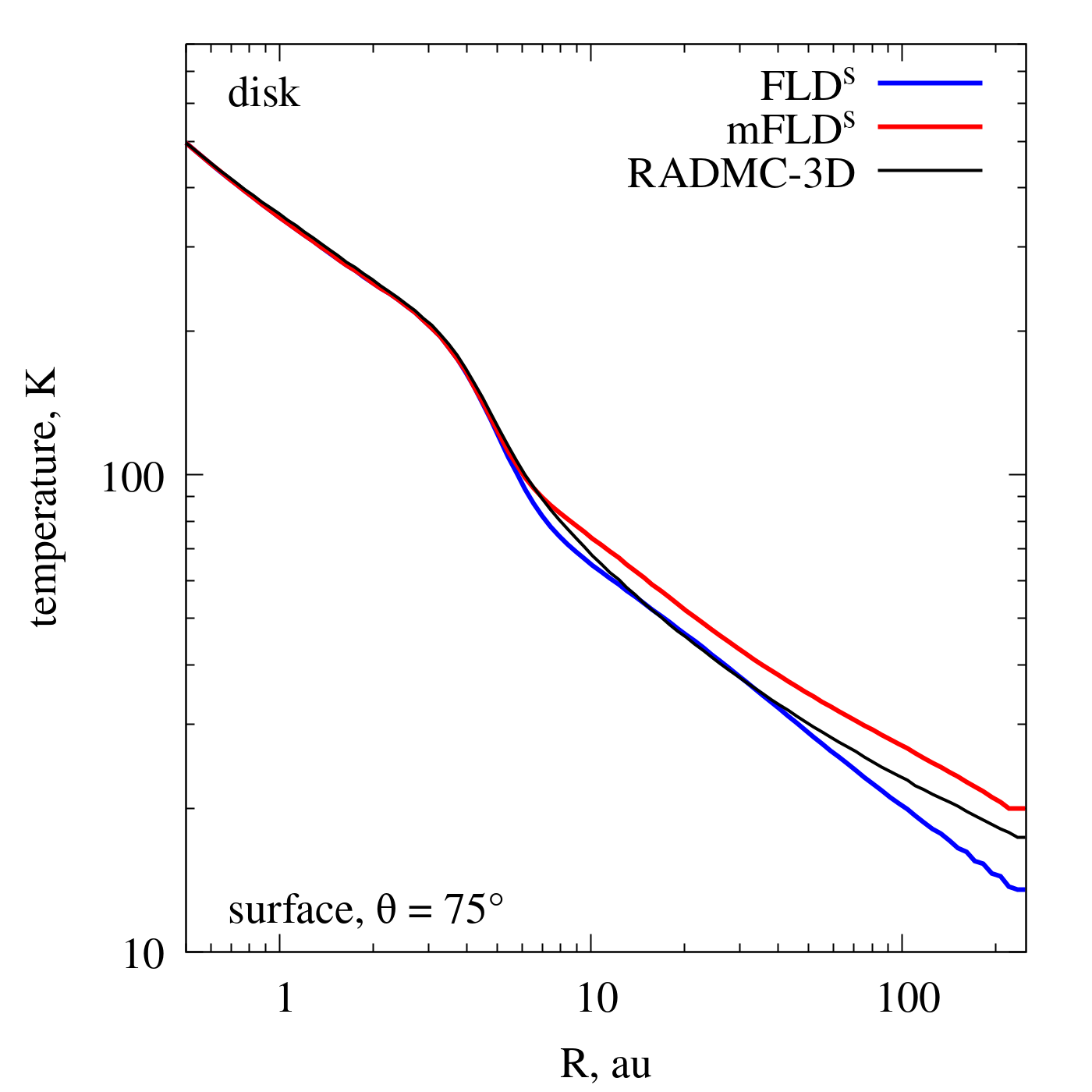}
\includegraphics[angle=0,width=0.23\textwidth]{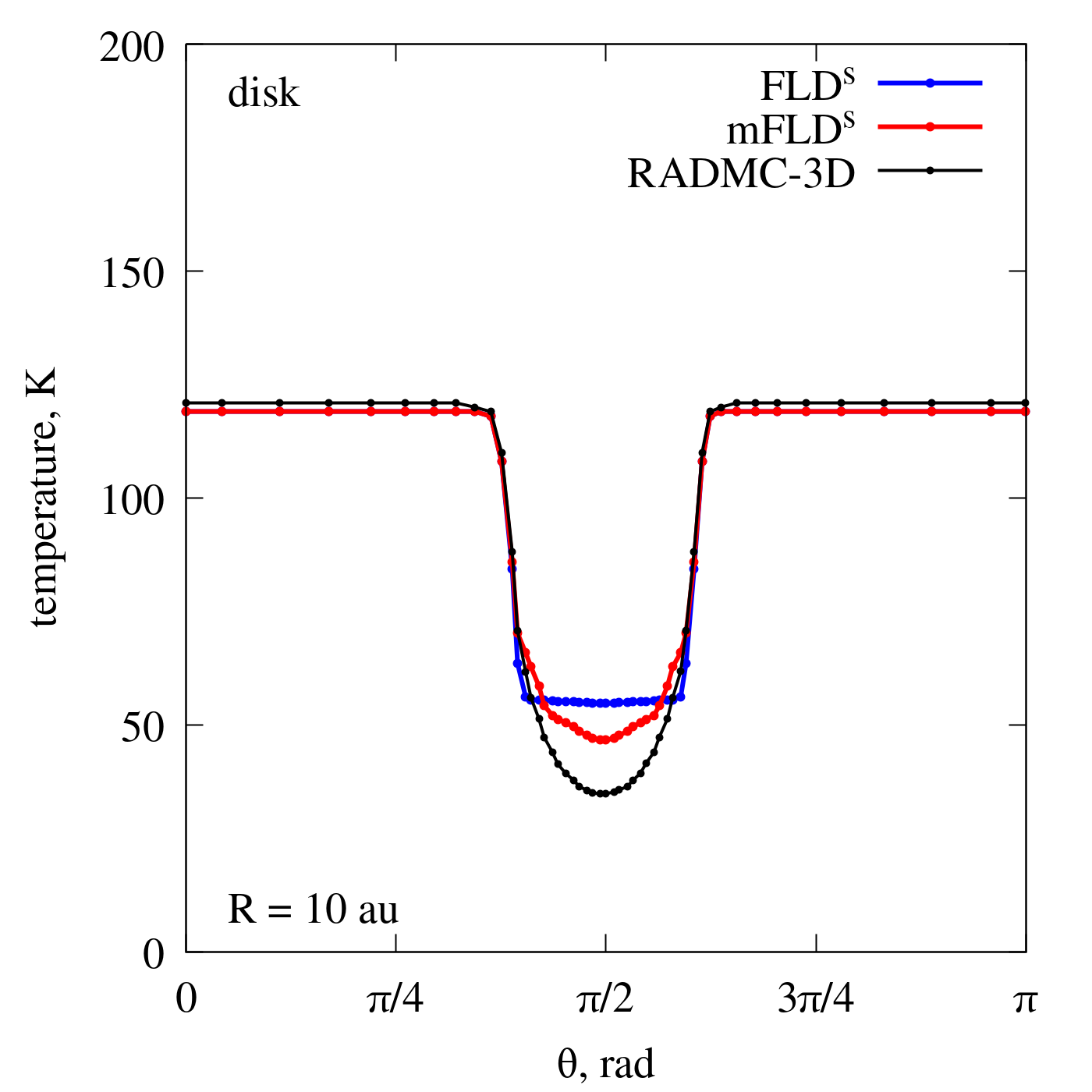}\\
\includegraphics[angle=0,width=0.23\textwidth]{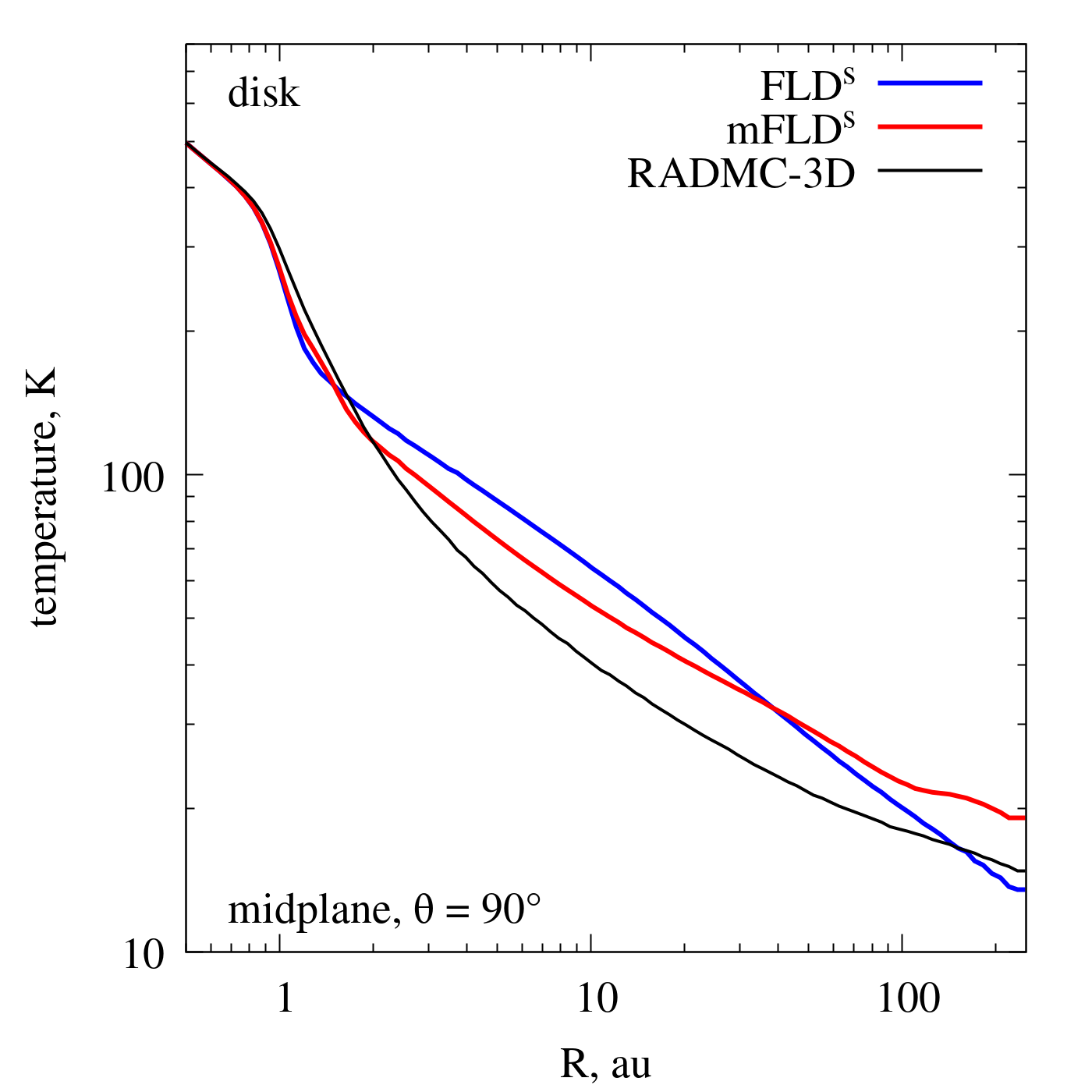}
\includegraphics[angle=0,width=0.23\textwidth]{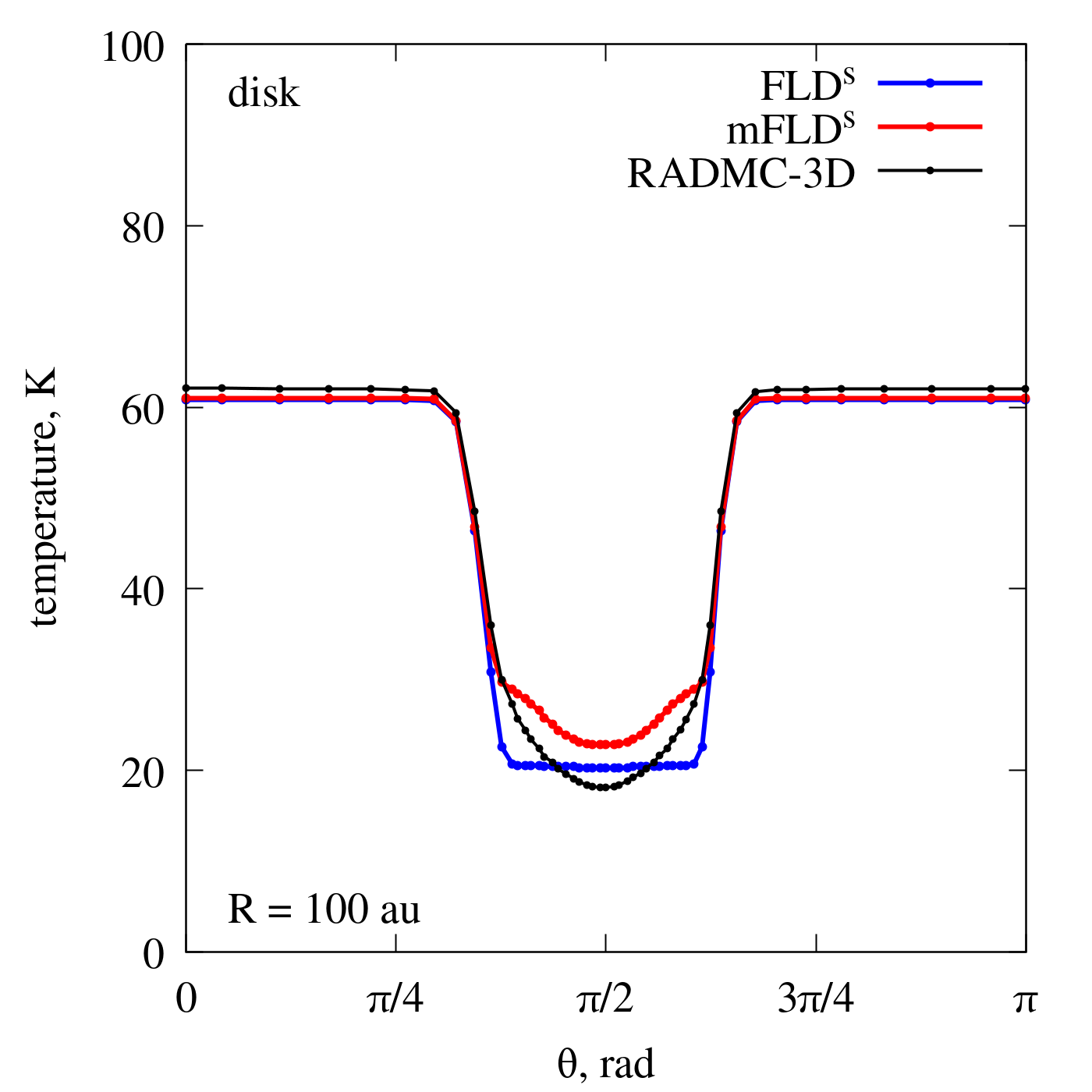}
\caption{Stationary temperature distributions for the gas-dust disk
model, obtained by the FLD$^{\rm s}$, mFLD$^{\rm s}$ methods, and the
RADMC-3D code. Left column: radial cuts along different polar angles.
Right column: angular cuts along different radial positions.}
\label{fig:compareA4}
\end{figure}

To describe the structure of a protoplanetary disk, we take the M3 model
from Paper~I. The central star parameters are the following:
$M_{\ast}=0.5\,M_\odot$, $L_{\ast}=0.7\,L_\odot$, $T_{\star}=3800$~K. The
radial profile of the surface density in the disk:
\begin{equation}
    \Sigma(r)=\Sigma_0 \left(\frac{r}{r_\text{2}}\right)^{-\gamma}
    \exp\left[
    -\left(\frac{r}{r_\text{2}}\right)^{2-\gamma}\right]\exp\left[
    -\left(\frac{r}{r_\text{1}}\right)^{\gamma-2}\right],
    \label{eq_surf}
\end{equation}
where $r$ is the distance from the polar axis, $\gamma=1$, the parameters
$r_\text{1}=10$ au and $r_\text{2}=70$ au determine the smoothing of the
distribution to the inner and outer boundaries of the grid, which are
chosen equal to $R_\text{in}=0.5$ au and $R_\text{out}=250$ au. The dust-to-gas mass ratio is assumed to be uniform over the disk and
equal to 0.01. The total mass of the disk is $M_\text{disk}=2.5\times
10^{-2} M_\odot$. The two-dimensional density distribution (see the left
upper panel in Fig.~\ref{fig:compareA3}) is calculated from $\Sigma(r)$
assuming vertical hydrostatic equilibrium at a fixed temperature
distribution $T(r) = 300 \left({r}/{1\,\mbox{au}}\right)^{-0.5}$ K.

\begin{figure}
\includegraphics[angle=0,width=0.23\textwidth]{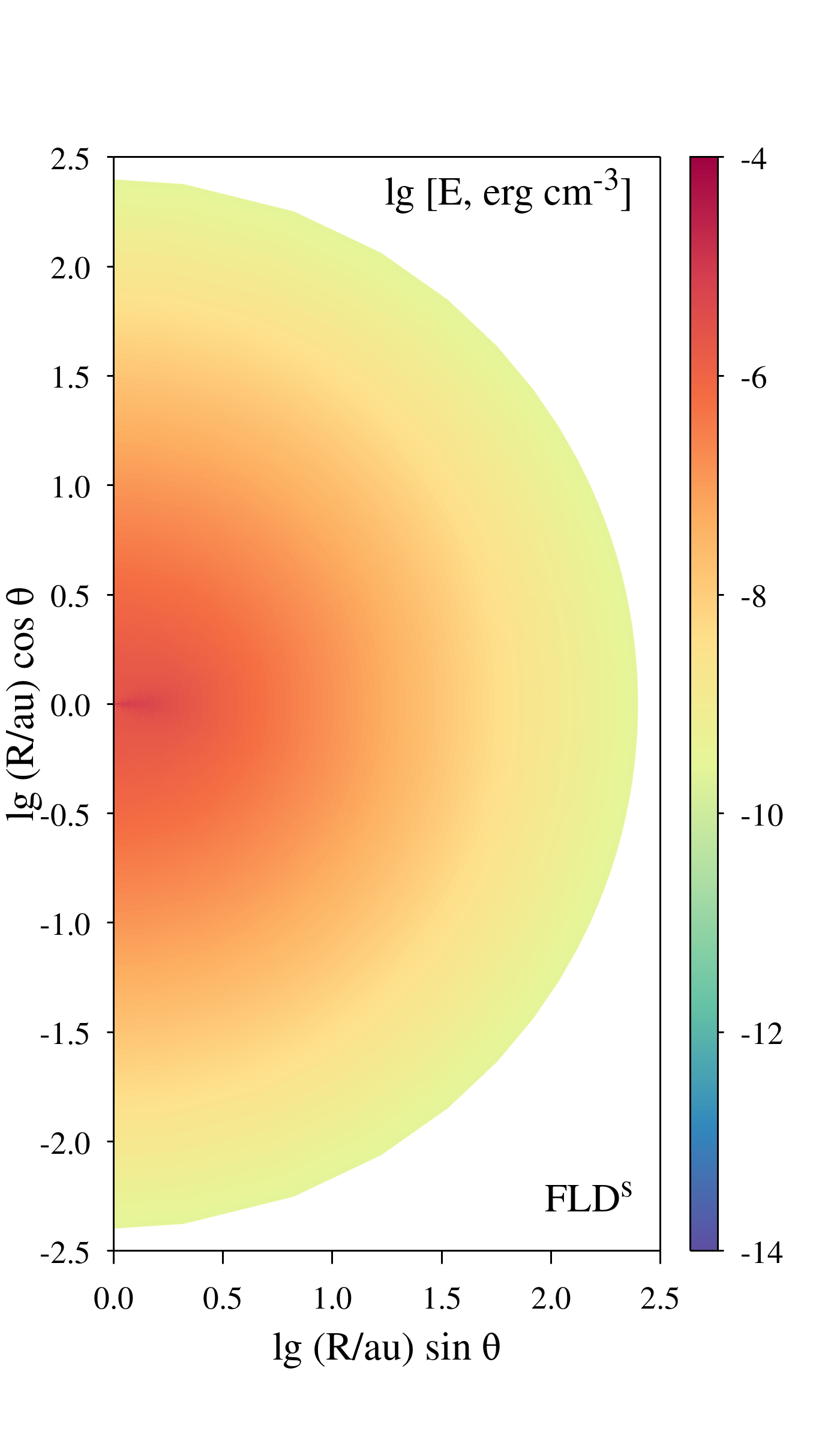}
\includegraphics[angle=0,width=0.23\textwidth]{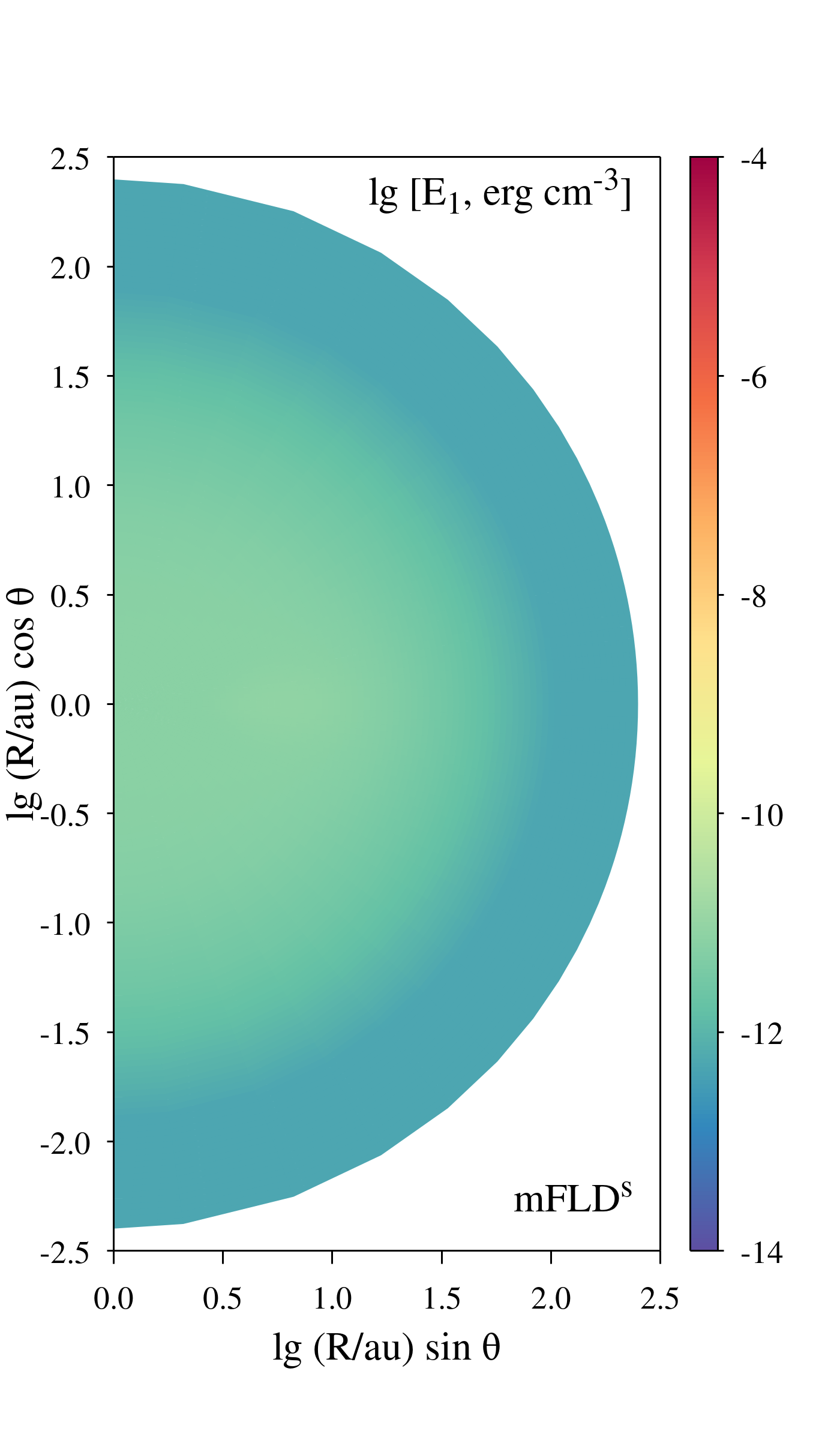}\\
\includegraphics[angle=0,width=0.23\textwidth]{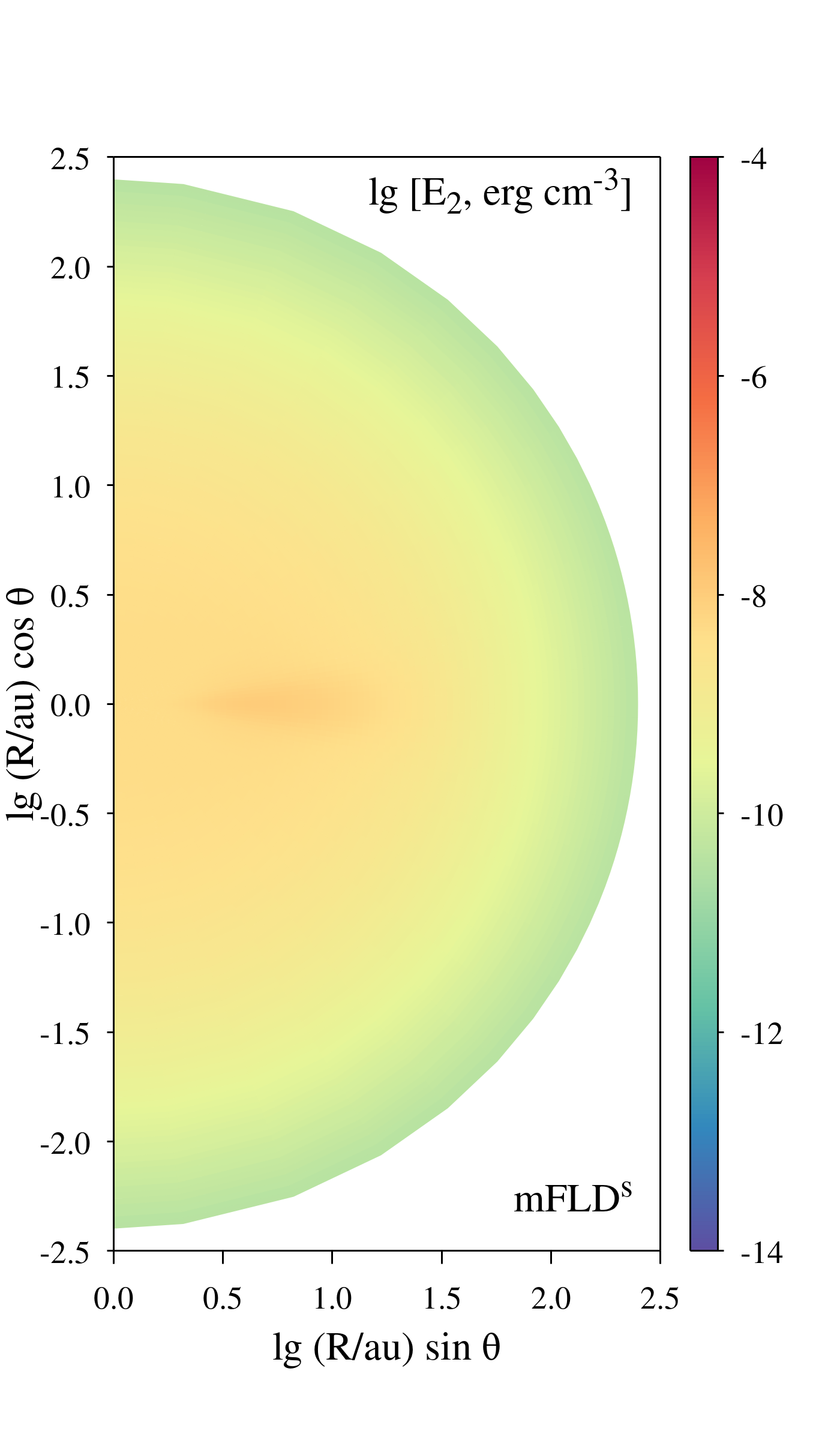}
\includegraphics[angle=0,width=0.23\textwidth]{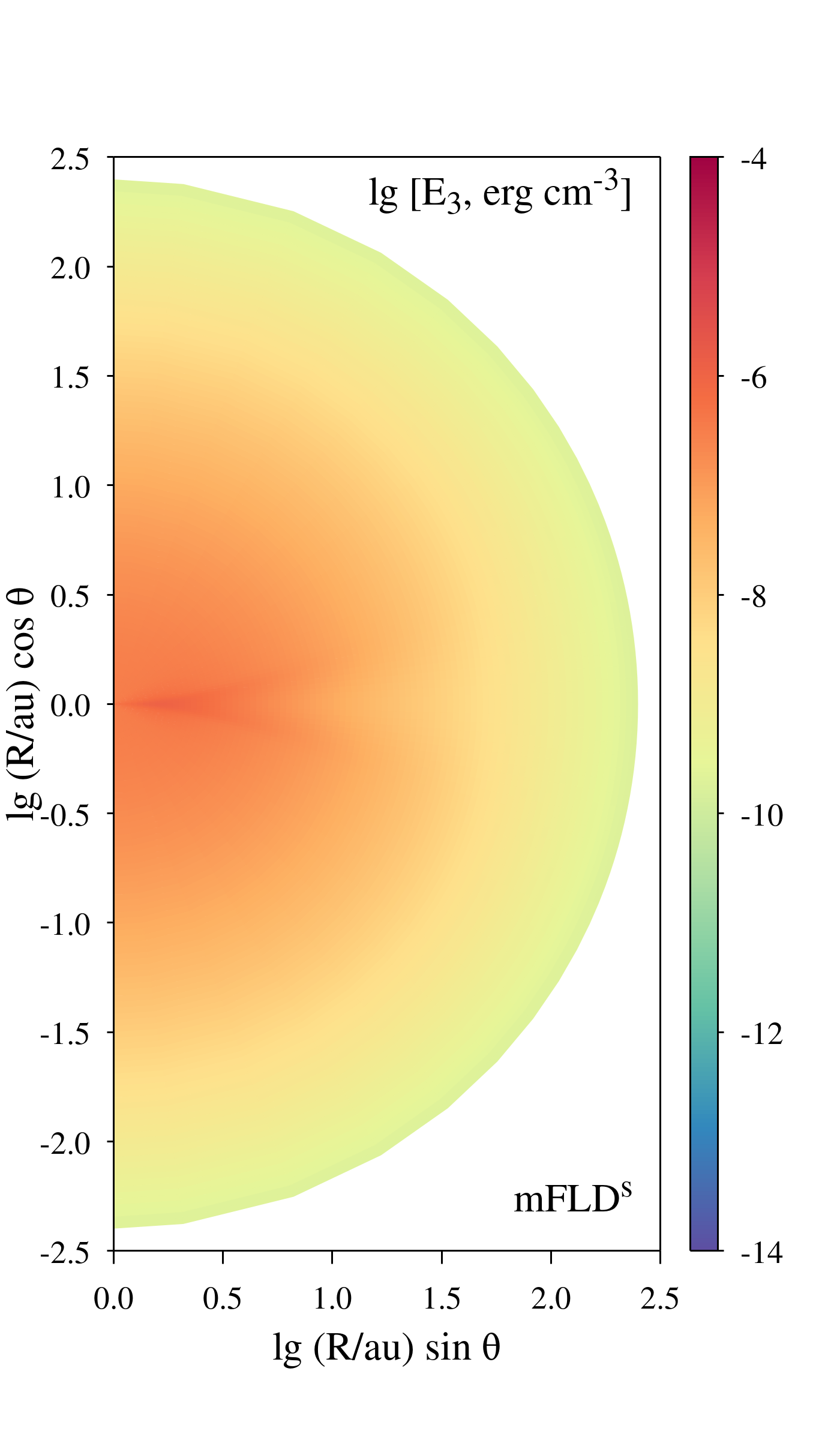}\\
\includegraphics[angle=0,width=0.23\textwidth]{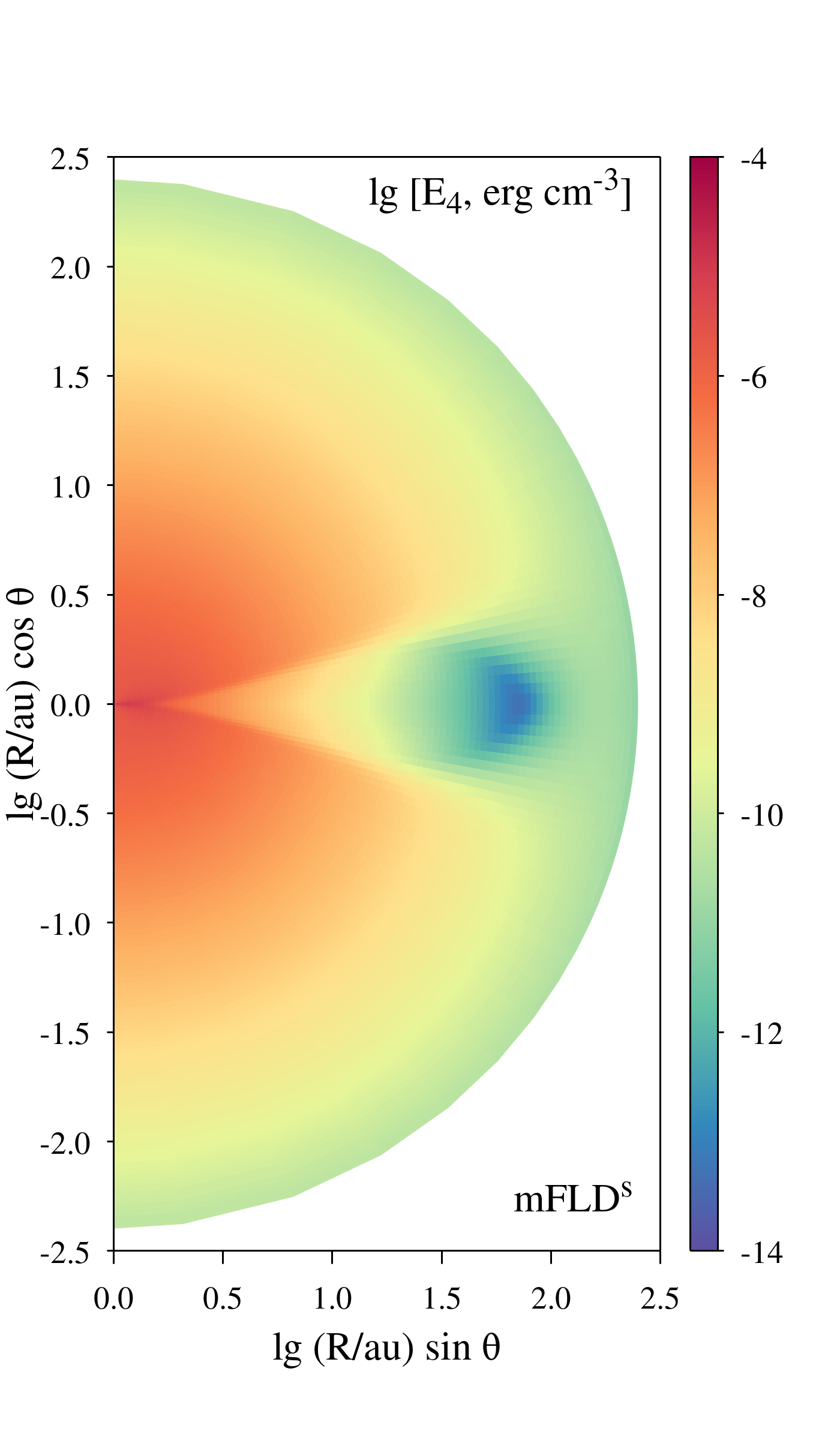}
\includegraphics[angle=0,width=0.23\textwidth]{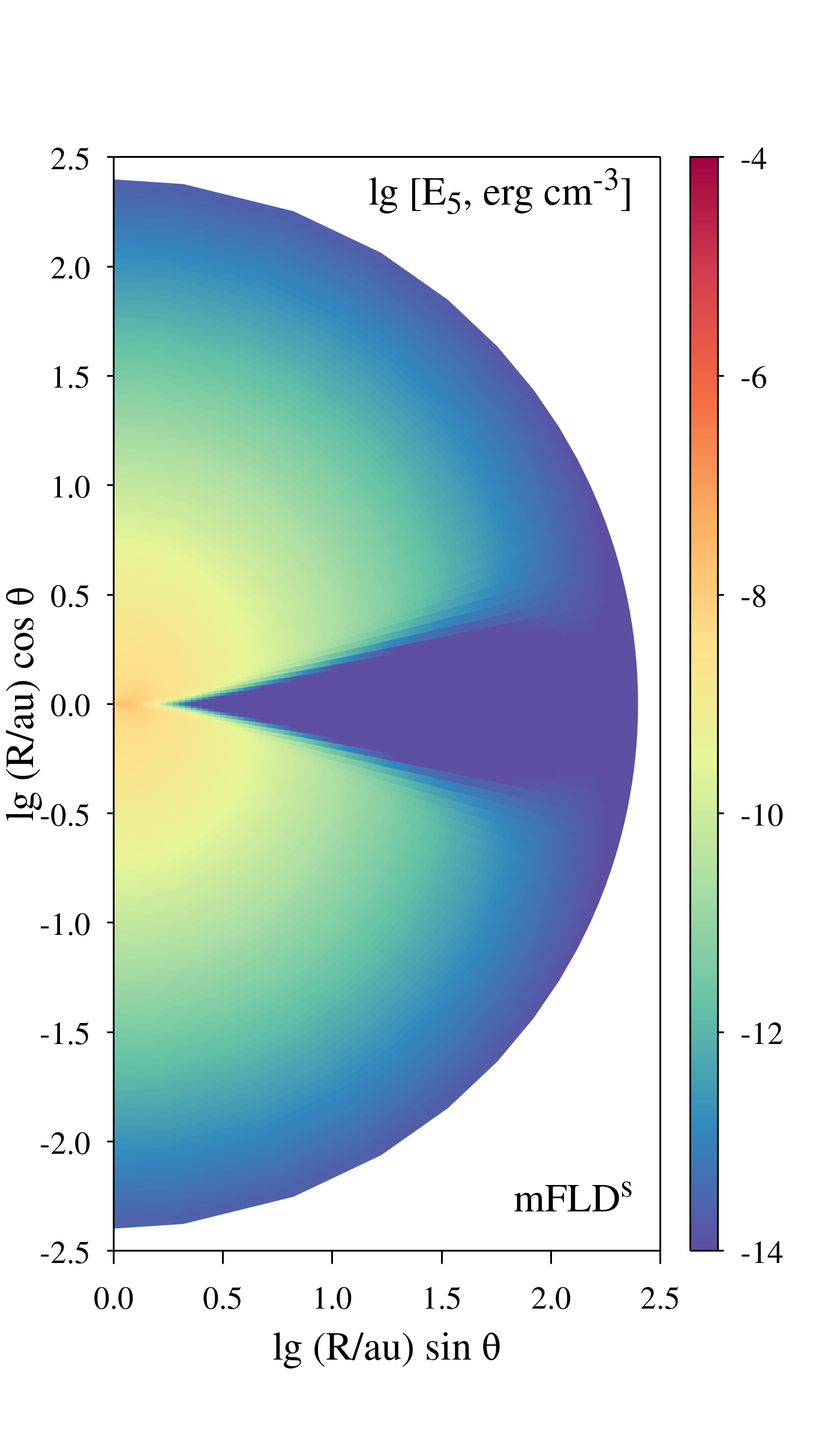}
\caption{Top left panel: stationary distribution of radiation energy
density for the protoplanetary disk model, obtained by the FLD$^{\rm s}$
method. The remaining panels: stationary distributions of radiation energy
density within the selected frequency intervals, obtained by the
mFLD$^{\rm s}$ method.}
\label{fig:compareA5}
\end{figure}

Let us analyze the results of calculating the stationary thermal
structure using the FLD$^{\rm s}$, mFLD$^{\rm s}$ methods, as well as the
RADMC-3D code. The solution for FLD$^{\rm s}$, mFLD$^{\rm s}$ is chosen
at the moment $10^4$~years, which is many times longer than the time to
reach a stationary state. Figure~\ref{fig:compareA3} shows the
two-dimensional temperature distributions in the polar section of the
disk, using for the abscissa and ordinate axes the coordinates
$\lg\left(R/\text{au}\right)\sin\theta$,
$\lg\left(R/\text{au}\right)\cos\theta$, which allow a more
detailed view of the inner regions of the disk.
Figure~\ref{fig:compareA4} shows the temperature cuts along radial
directions and along the angle $\theta$ at various distances from the
star.

The temperature distributions in the disk envelope in all three methods
are {monotonic and} identical, which is natural, since the envelope
is heated by direct stellar radiation, and the details of thermal
radiation transfer are not important here, since the envelope is
transparent to it.
The temperature distribution in the disk midplane,
obtained by the FLD$^{\rm s}$ method (see the bottom left panel
of Fig.~\ref{fig:compareA4}), can be roughly divided into three regions
between which the slope of the distribution changes.
Note that when using the surface density
profile, eq.~\eqref{eq_surf}, the gas density in the disk miplane first increases
outward from the star, reaching a maximum around 4~au, and then decreases.
In the inner region ($R<0.9$ au), the medium is transparent to direct stellar
radiation, and the temperature here is determined by the dilution of stellar
radiation. In the middle region ($R=0.9-1.3$ au), the miplane regions become
opaque to direct stellar radiation but remain transparent to the disk's own
thermal radiation. In this region, the thermal radiation freely escapes through
the inner boundary of the disk, cooling the medium and resulting in a steeper
temperature drop with radius.
In the outer region ($R>1.3$ au), the optical depth for the escape of thermal
radiation becomes significant, the radiation becomes "trapped" and propagates
in diffusion mode, leading to a change in the slope of the temperature
distribution. A similar pattern is observed for the temperature distribution
along the $\theta=75^{\circ}$ direction (middle left panel of Fig.~\ref{fig:compareA4}),
but the breaks in the temperature distributions are shifted to greater distances
from the star.

As noted in Paper~I, the midplane temperature distribution between 2
and 150 au, obtained by the FLD$^{\rm s}$ method, is significantly higher
than the reference distribution and follows the scaling $T\propto
R^{-1/2}$, which does not agree with the variable slope of the exact
distribution. The midplane temperature distribution from the mFLD$^{\rm
s}$ method has a variable slope and is closer to the reference one. From
the comparison of the cuts along the polar angle, it is also seen that
the morphology of the mFLD$^{\rm s}$ distribution repeats the morphology
of the profile from RADMC-3D, while the angular temperature dependence
from FLD$^{\rm s}$ shows an isothermal plateau in the vicinity of the
midplane.

\begin{figure*}
\includegraphics[angle=0,width=0.27\textwidth]{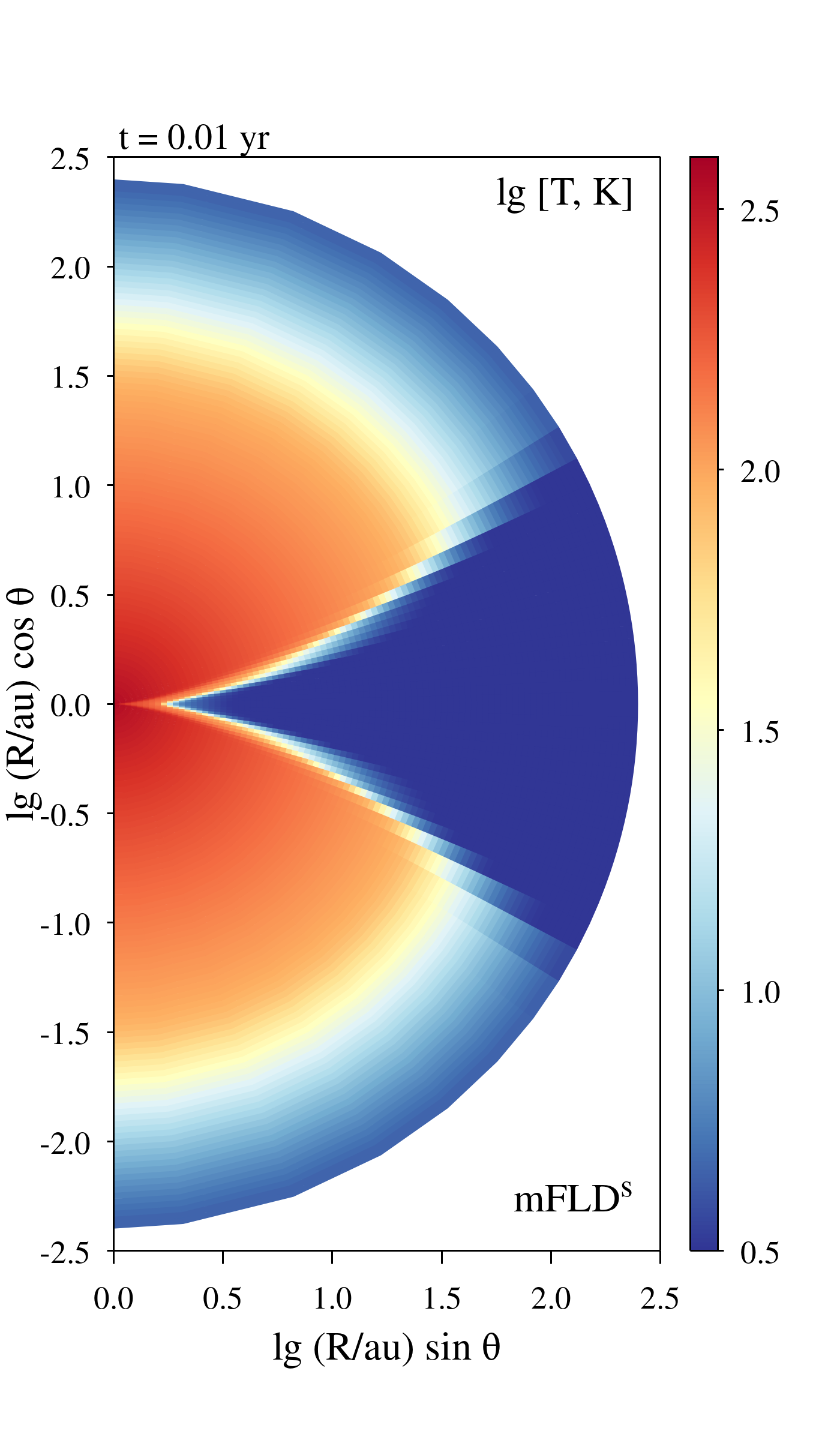}
\includegraphics[angle=0,width=0.27\textwidth]{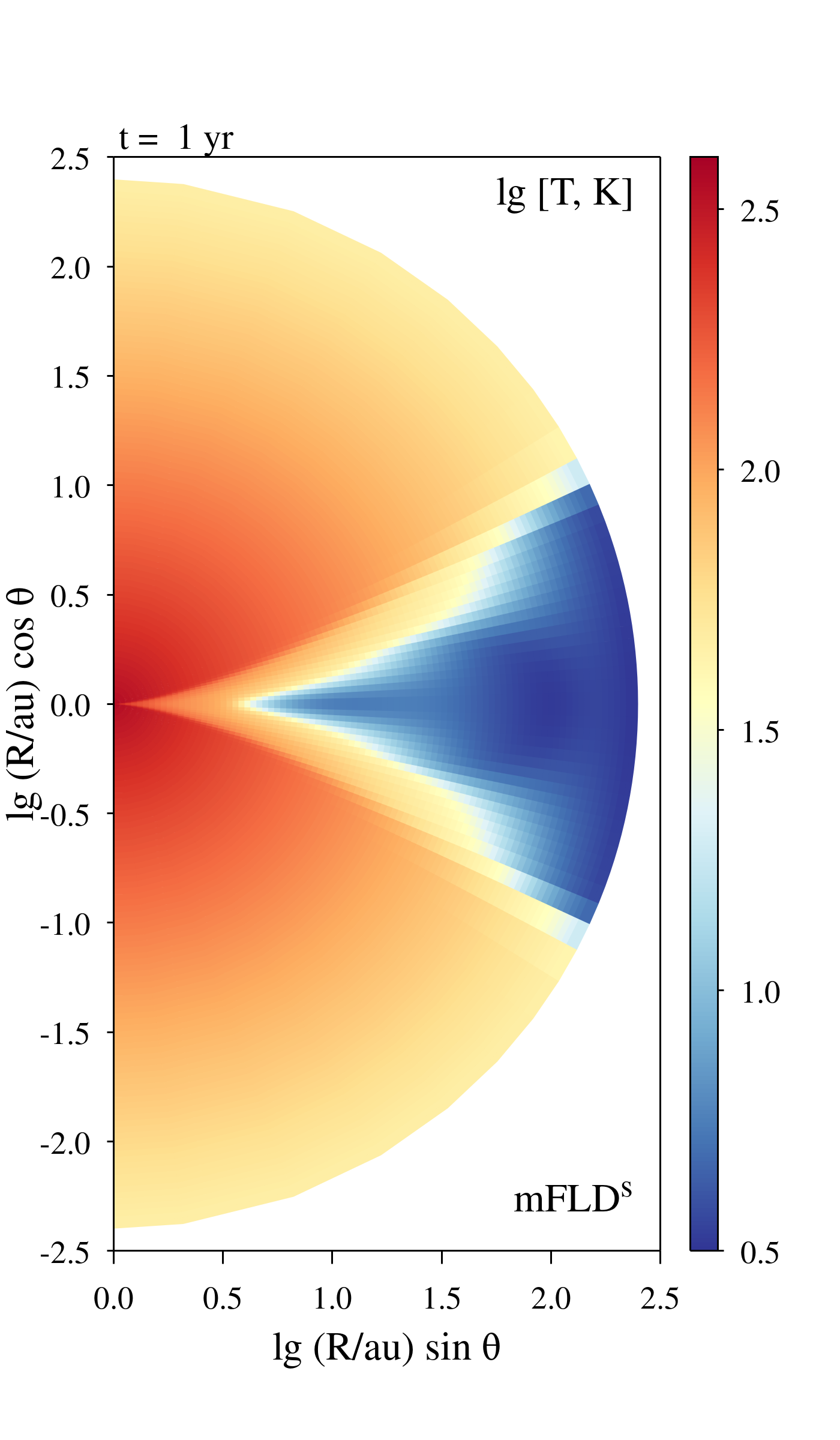}
\includegraphics[angle=0,width=0.27\textwidth]{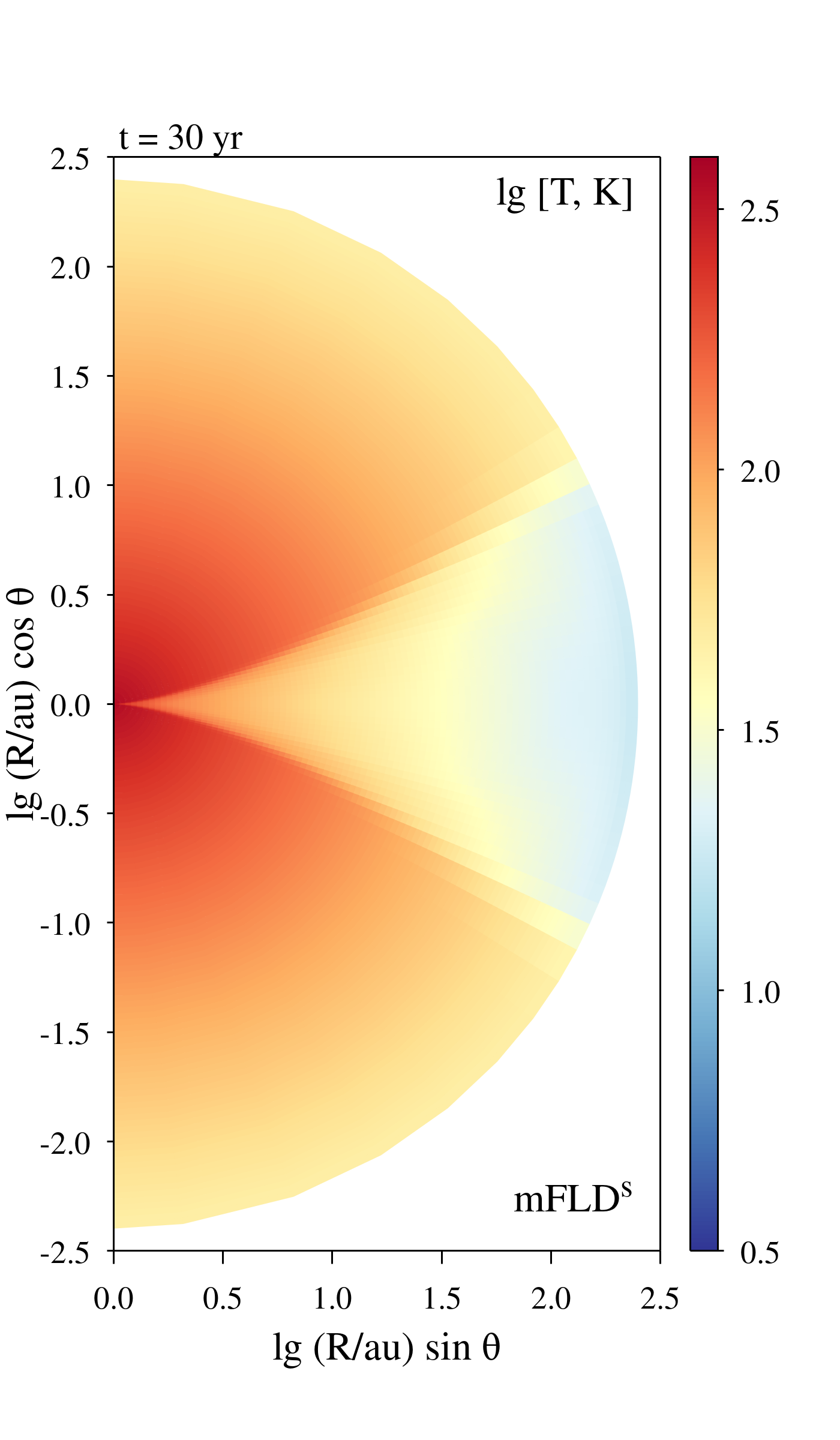} \\
\includegraphics[angle=0,width=0.36\textwidth]{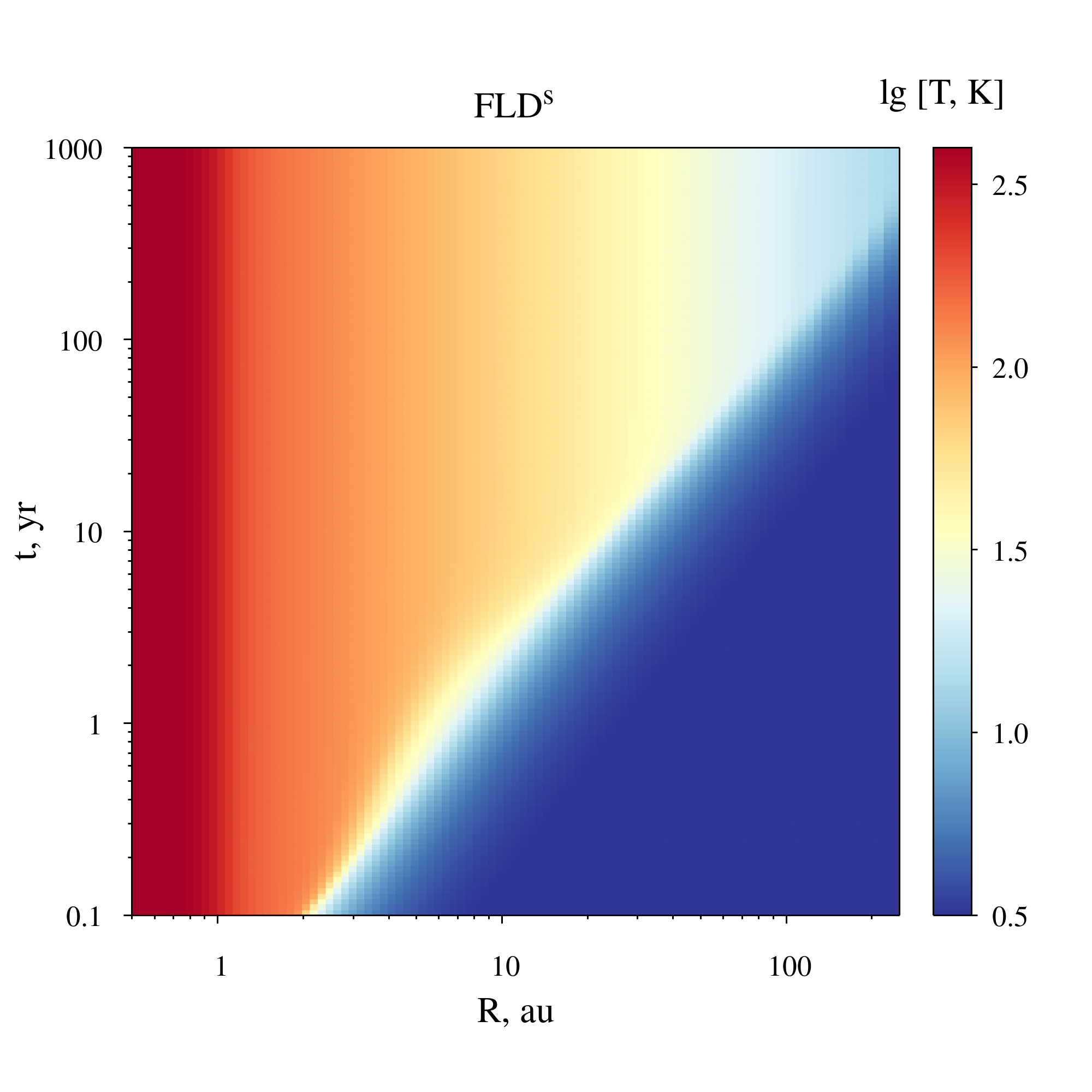}
\includegraphics[angle=0,width=0.36\textwidth]{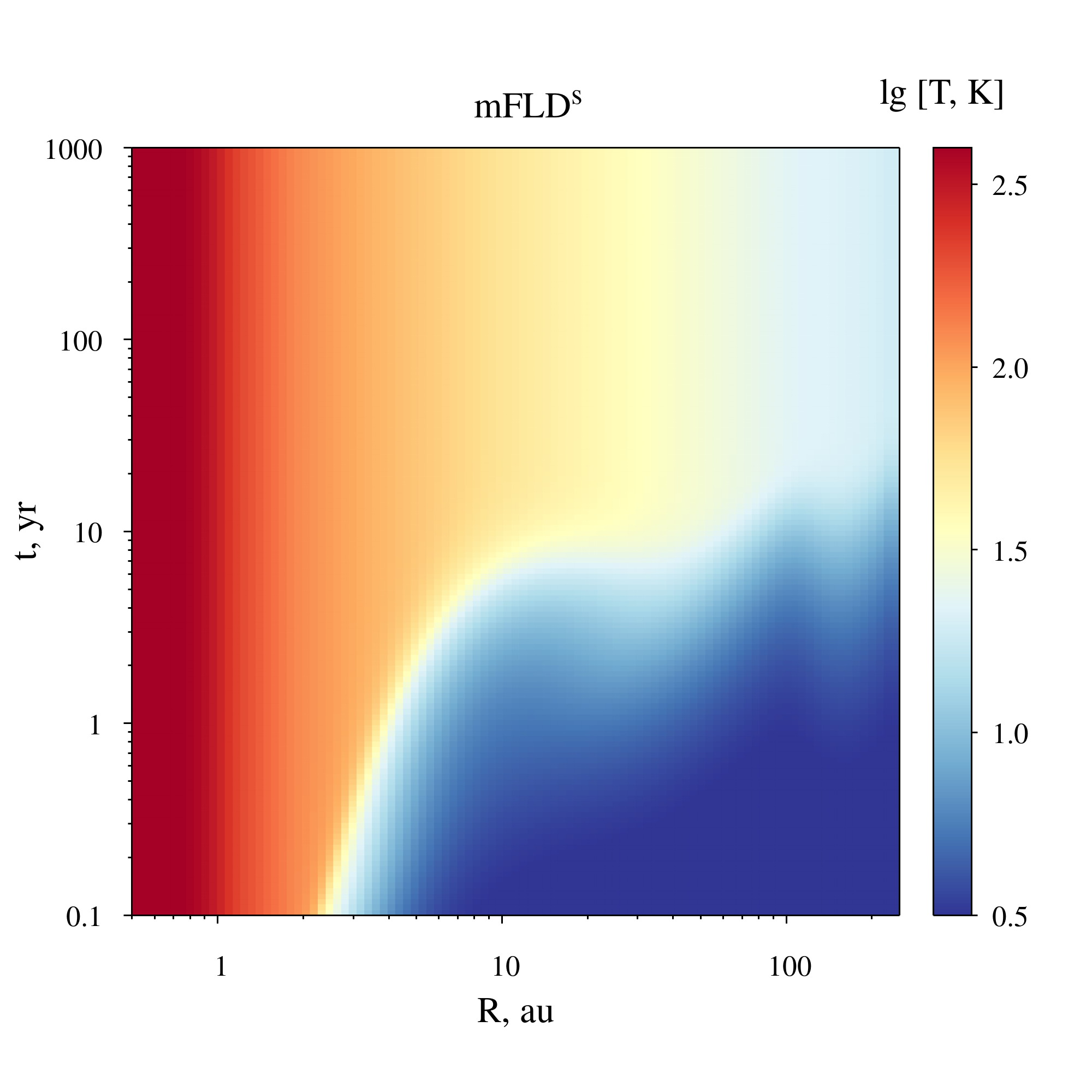}
\caption{Top row: temperature distributions in the polar section of the
disk at different times, obtained using mFLD$^{\rm s}$. The times are
indicated above the panels. Bottom panel: evolution of the midplane
temperature distribution to a stationary state. Left: FLD$^{\rm s}$,
right: mFLD$^{\rm s}$. The color scale boundaries differ from those in
Fig.~\ref{fig:compareA3}.}
\label{fig:compareA6}
\end{figure*}

The reason for the differences between the results of FLD$^{\rm s}$ and
mFLD$^{\rm s}$ can be understood by comparing the stationary
distributions of radiation energy density obtained by these methods, see
Fig.~\ref{fig:compareA5}. The distribution of radiation energy density for
FLD$^{\rm s}$ is close to spherically symmetric. This is a consequence of
the fact that thermal radiation in this case is "locked" in one spectral
channel --- in an optically thick medium, it is absorbed by matter,
coming into thermal equilibrium with it, re-emitted, and ultimately can
leave the medium only due to its own diffusion, which leads to a high
degree of symmetry.
In contrast, in the mFLD$^{\rm s}$ method, the radiation energy is not only
subject to diffusion but can transfer between spectral channels as a
result of absorption and re-emission.
The morphology of the distributions of $E_{m}$ (energy density within the
$m$-th frequency channel) in the mFLD$^{\rm s}$ calculation changes from
almost homogeneous to strongly asymmetric when moving from the first to
the last channel. In the first and second channels, the disk is almost
transparent to its own thermal radiation, so diffusion is effective, and
the distributions of $E_1$ and $E_2$ are close to homogeneous. In the
fifth frequency channel, the energy density inside the disk is extremely
low, since stellar radiation from the near-star regions does not
penetrate here, and the temperature of the medium itself is too low to
generate its own. The highest energetics is observed in the third and fourth
frequency channels, where the energy of the own thermal radiation is
predominantly concentrated. In channels 3 and 4, the asymmetry of the
radiation energy density distributions is also pronounced. Thus, the
division of the frequency range into frequency channels within the FLD
approach allows longer-wavelength thermal radiation to propagate more
freely through the disk. On the one hand, this allows radiation to more
easily penetrate into the disk and heat it, and on the other hand, to
more freely leave the disk and cool it, depending on the conditions.

In general, it can be concluded that the use of mFLD$^{\rm s}$ eliminates
a number of problems of the FLD$^{\rm s}$ method in modeling the thermal
structure of a protoplanetary disk, but some quantitative differences
with the exact results remain.

\subsection{Relaxation of the Temperature Distribution to a Stationary State}

Let us compare the FLD$^{\rm s}$ and mFLD$^{\rm s}$ methods in a
case that assumes significant non-stationarity of the thermal
structure. Comparison with the RADMC-3D code was not carried out, since
it's current version does not support non-stationary radiation transfer.
The initial temperature of the disk is artificially set to 2.73 K. The
disk is heated by radiation from a star with the parameters from the
previous section. The upper panel of Fig.~\ref{fig:compareA6} shows the
temperature distributions in the polar section of the disk at different
times, calculated using mFLD$^{\rm s}$. The disk gradually heats up from
the upper layers towards midplane, with the heating zone shifting outward over
time. Fig.~\ref{fig:compareA6} (lower panels) also shows the complete
evolution of the midplane temperature distribution of the disk to a
stationary state, calculated by the FLD$^{\rm s}$ and mFLD$^{\rm s}$
methods. It can be seen that in the FLD$^{\rm s}$ approximation, the
outer parts of the disk reach a stationary state in 600 years, while the
characteristic time monotonically increases from the inner regions to the
outer ones. In the mFLD$^{\rm s}$ approximation, the outer layers reach
equilibrium much faster --- in 30 years, while the characteristic thermal
time changes with distance in a more complex way than in the FLD$^{\rm
s}$ model. The faster heating of the disk in the mFLD$^{\rm s}$
approximation is obviously related to the accelerated diffusion of
long-wavelength thermal radiation. The significant difference in thermal
times between FLD$^{\rm s}$ and mFLD$^{\rm s}$ leads to the conclusion
that caution should be exercised when using the FLD$^{\rm s}$ method for
modeling non-stationary processes in the disk (such as luminosity flares
and various instabilities), where the ratios of dynamic and thermal times
are important.

\subsection{Comparison of Methods in the Gray Approximation}

\begin{figure}
\includegraphics[angle=0,width=0.23\textwidth]{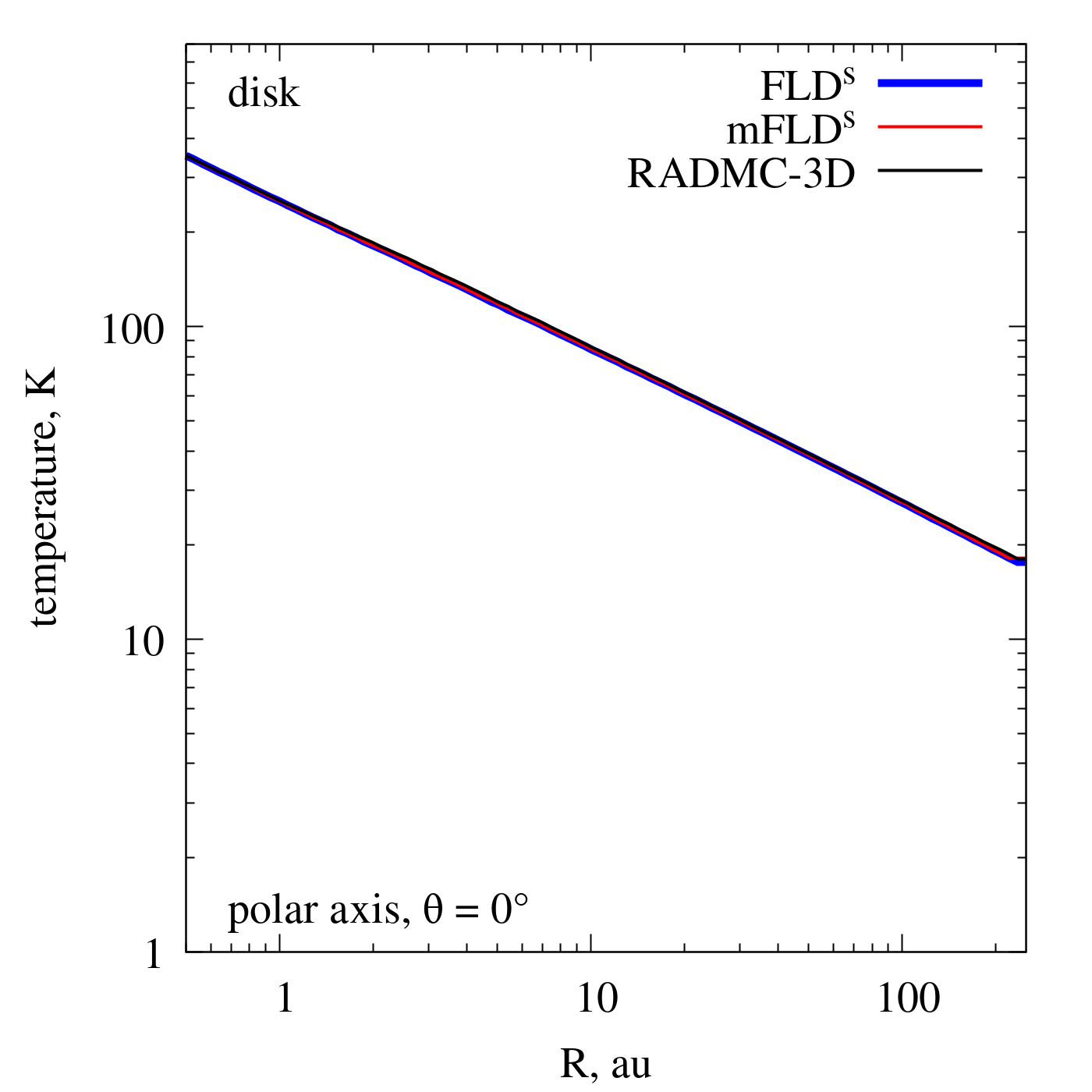}
\includegraphics[angle=0,width=0.23\textwidth]{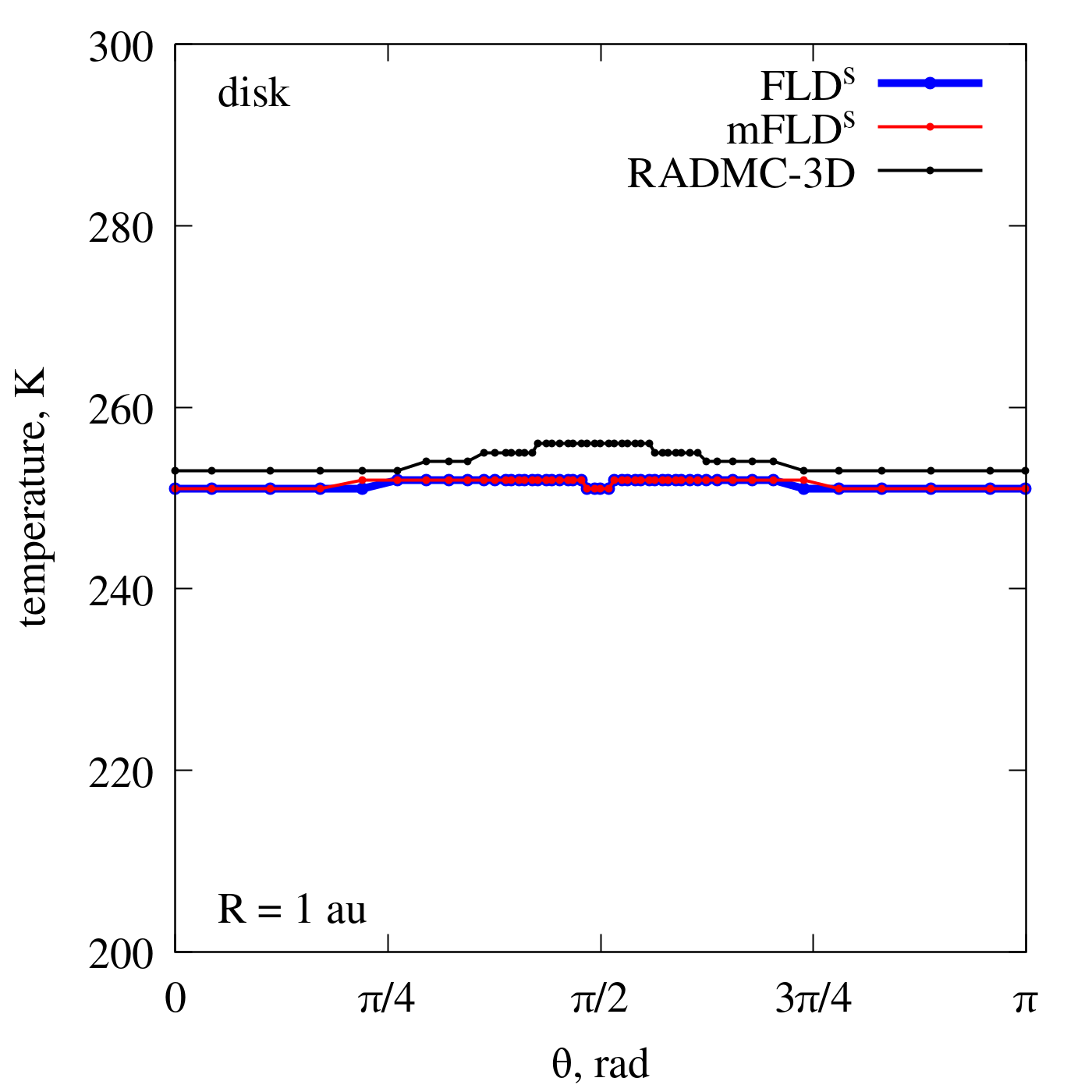}\\
\includegraphics[angle=0,width=0.23\textwidth]{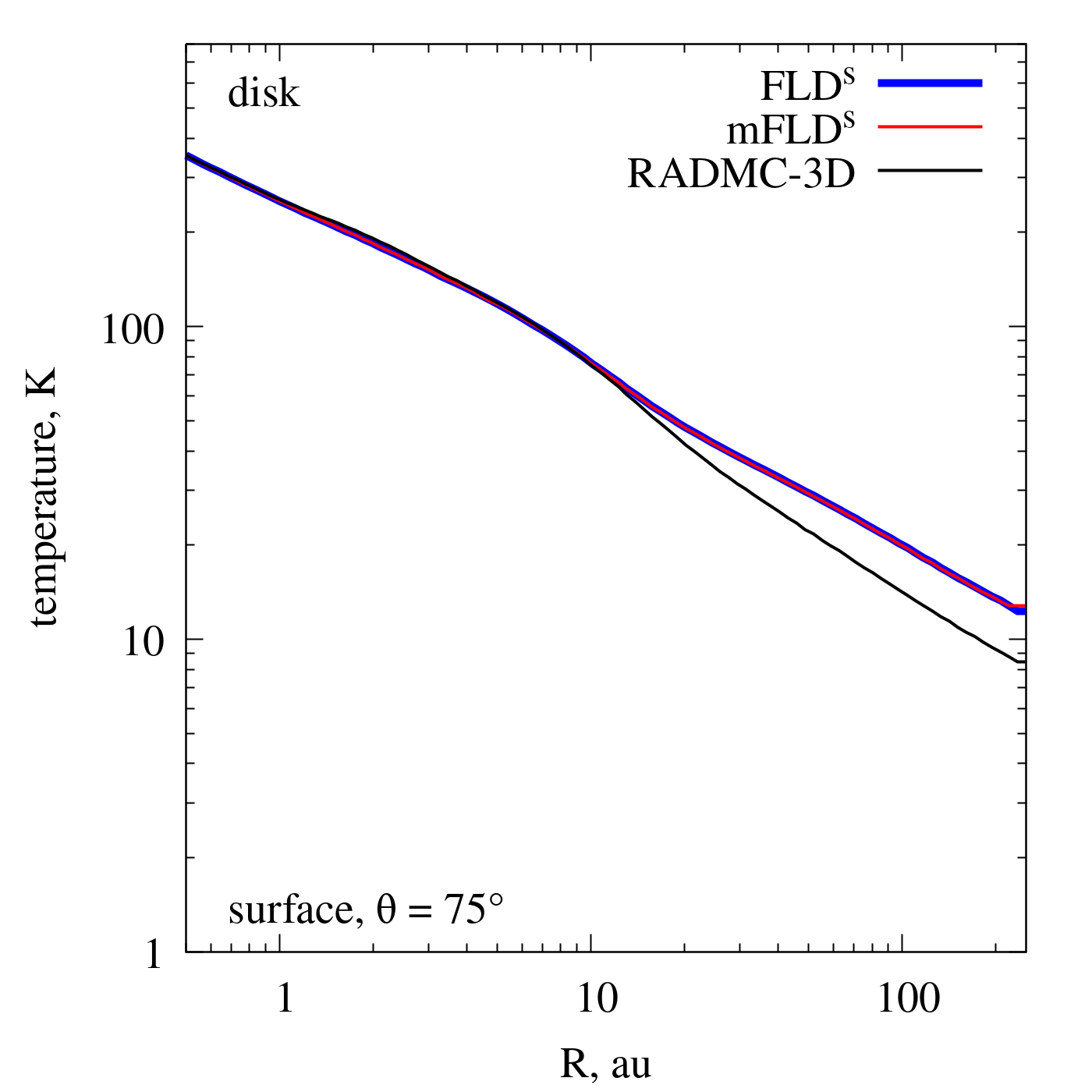}
\includegraphics[angle=0,width=0.23\textwidth]{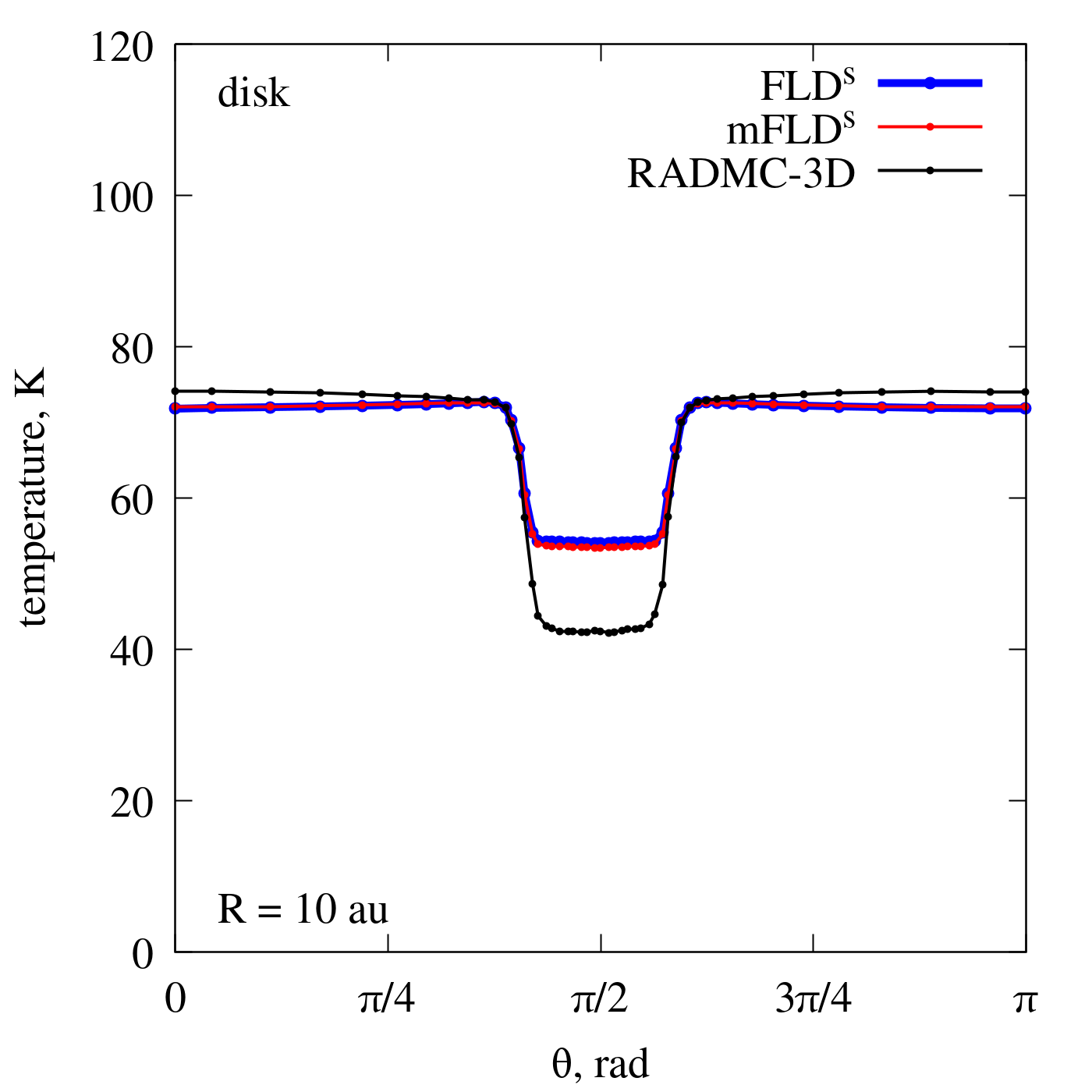}\\
\includegraphics[angle=0,width=0.23\textwidth]{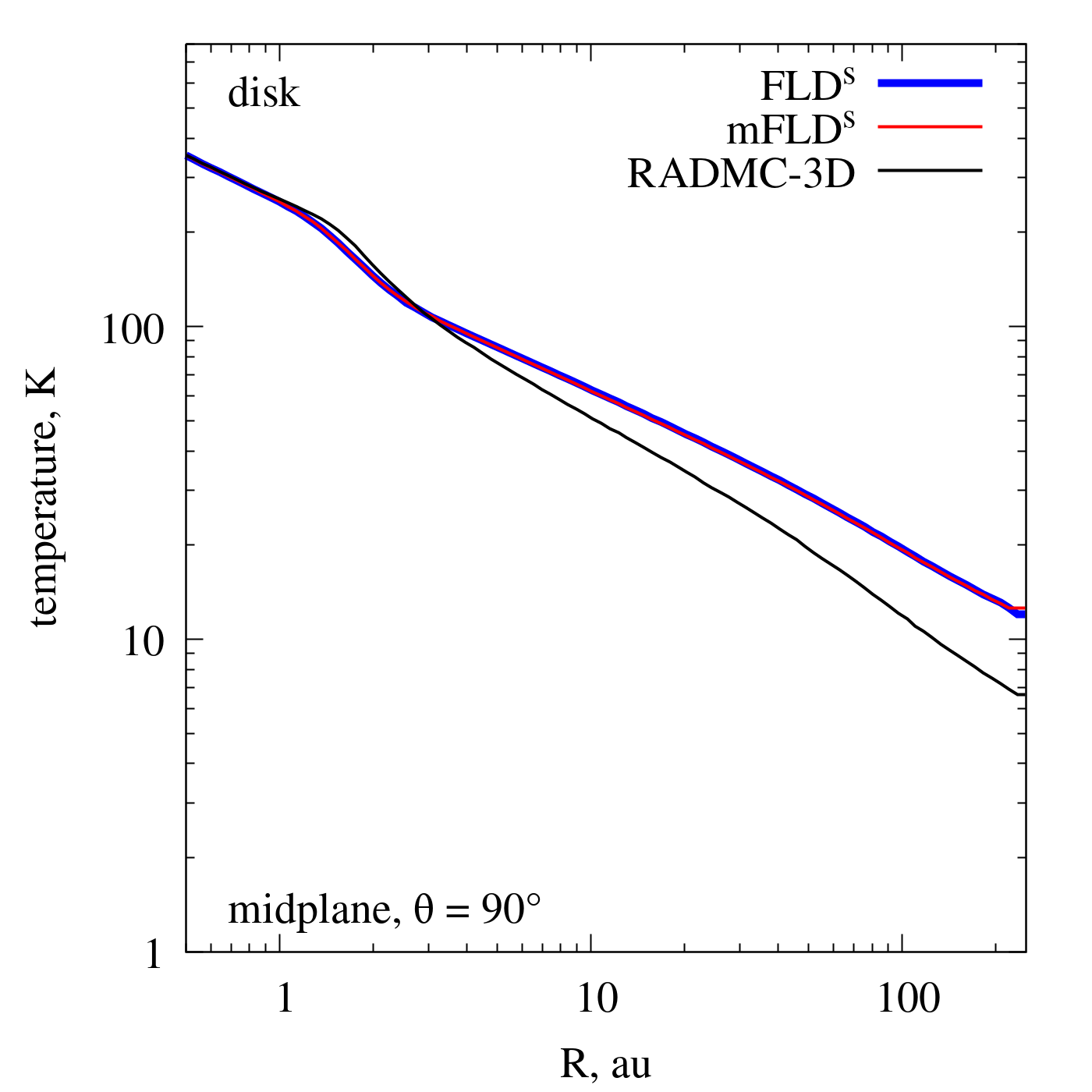}
\includegraphics[angle=0,width=0.23\textwidth]{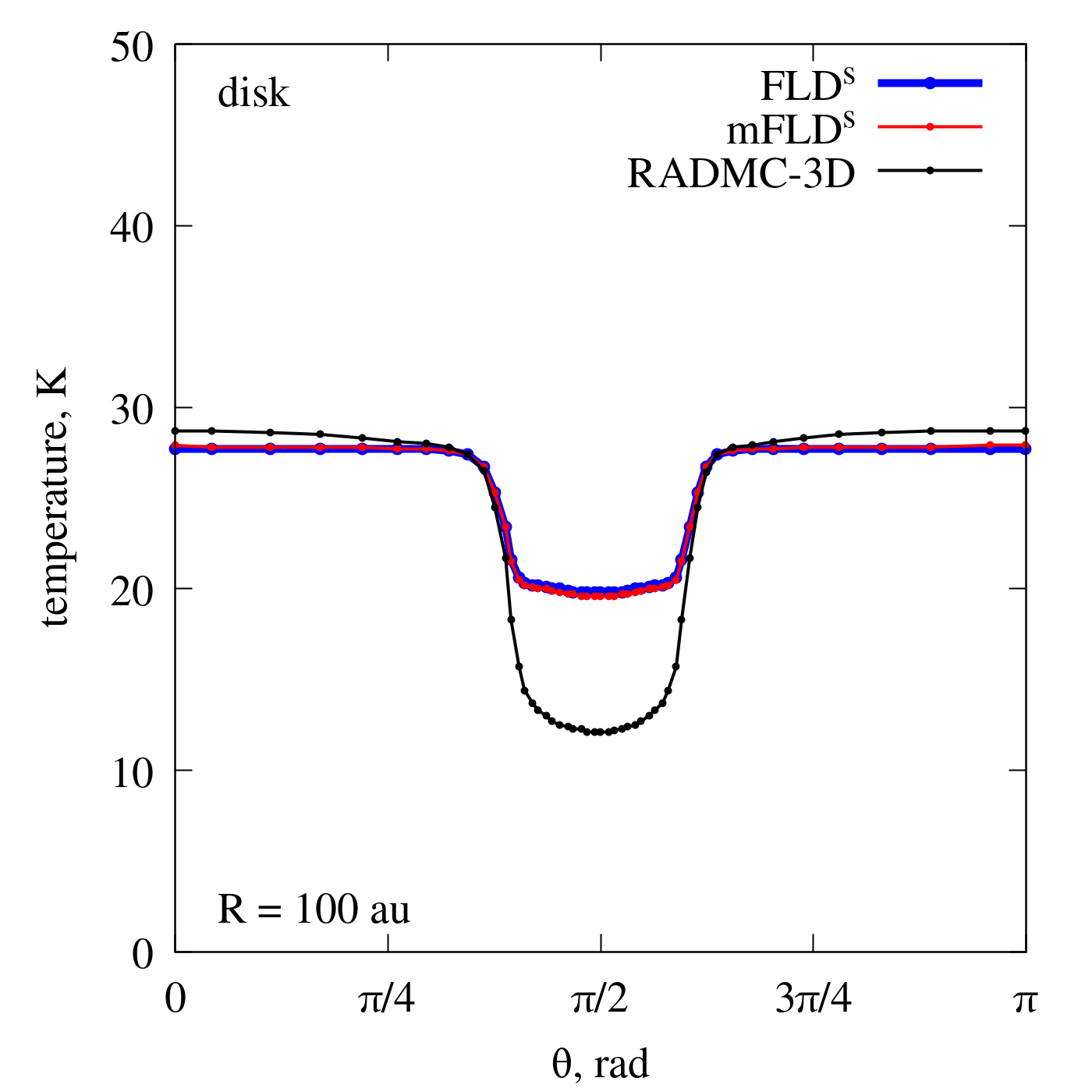}
\caption{Stationary temperature distributions for the gas-dust disk model
with frequency-independent opacities, obtained by the FLD$^{\rm s}$,
mFLD$^{\rm s}$ methods, and the RADMC-3D code. Left column: radial cuts along
different polar angles. Right column: angular cuts along different
radial positions.}
\label{fig:compareA7}
\end{figure}

Figure~\ref{fig:compareA7} shows the temperature distributions for the
gas-dust disk model, calculated under the assumption that the absorption
coefficient does not depend on frequency and is equal to $\kappa_{\nu}=1$
cm$^2$ g$^{-1}$ (dust). The distributions obtained by the FLD$^{\rm s}$
and mFLD$^{\rm s}$ methods, as expected, coincide. However, the results
obtained within the FLD approximations for the disk model differ from the
reference solution obtained by the RADMC-3D code. The midplane
temperature in the simulations with the RADMC-3D code decreases more rapidly with
distance than in the calculations using the FLD$^{\rm s}$ and mFLD$^{\rm
s}$ methods. This, in our notion, is related to the diffusive nature of
the FLD approximation and the isotropy of the diffusion coefficient (flux
limiter) implementing this approximation.

\subsection{Method Characteristics Depending on the Number of Frequency Channels}

\begin{figure*}
\includegraphics[angle=0,width=0.95\textwidth]{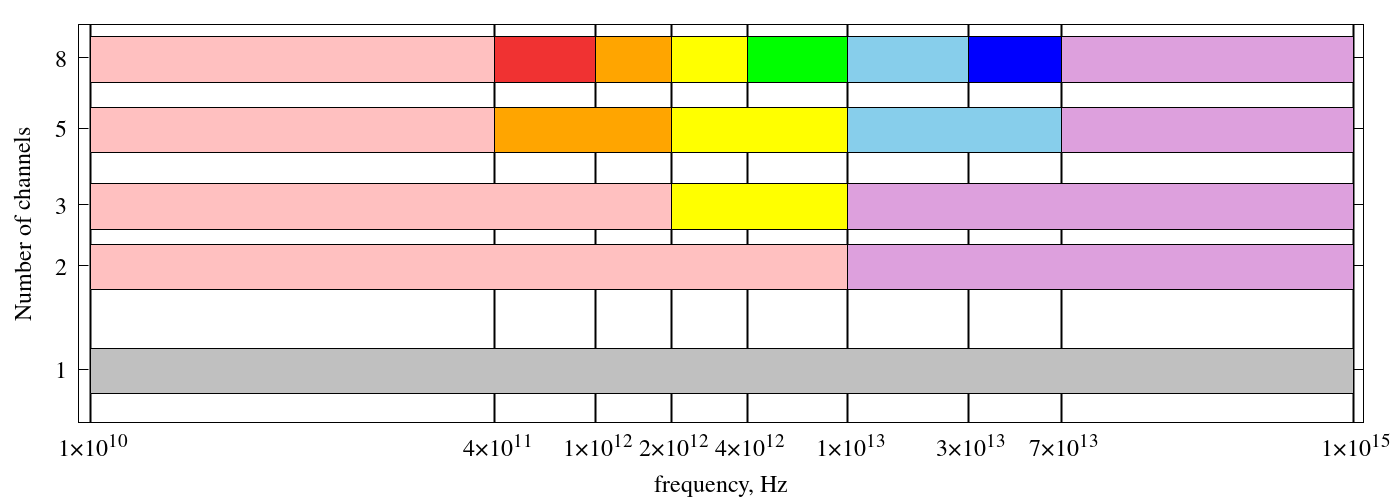} \\
\includegraphics[angle=0,width=0.48\textwidth]{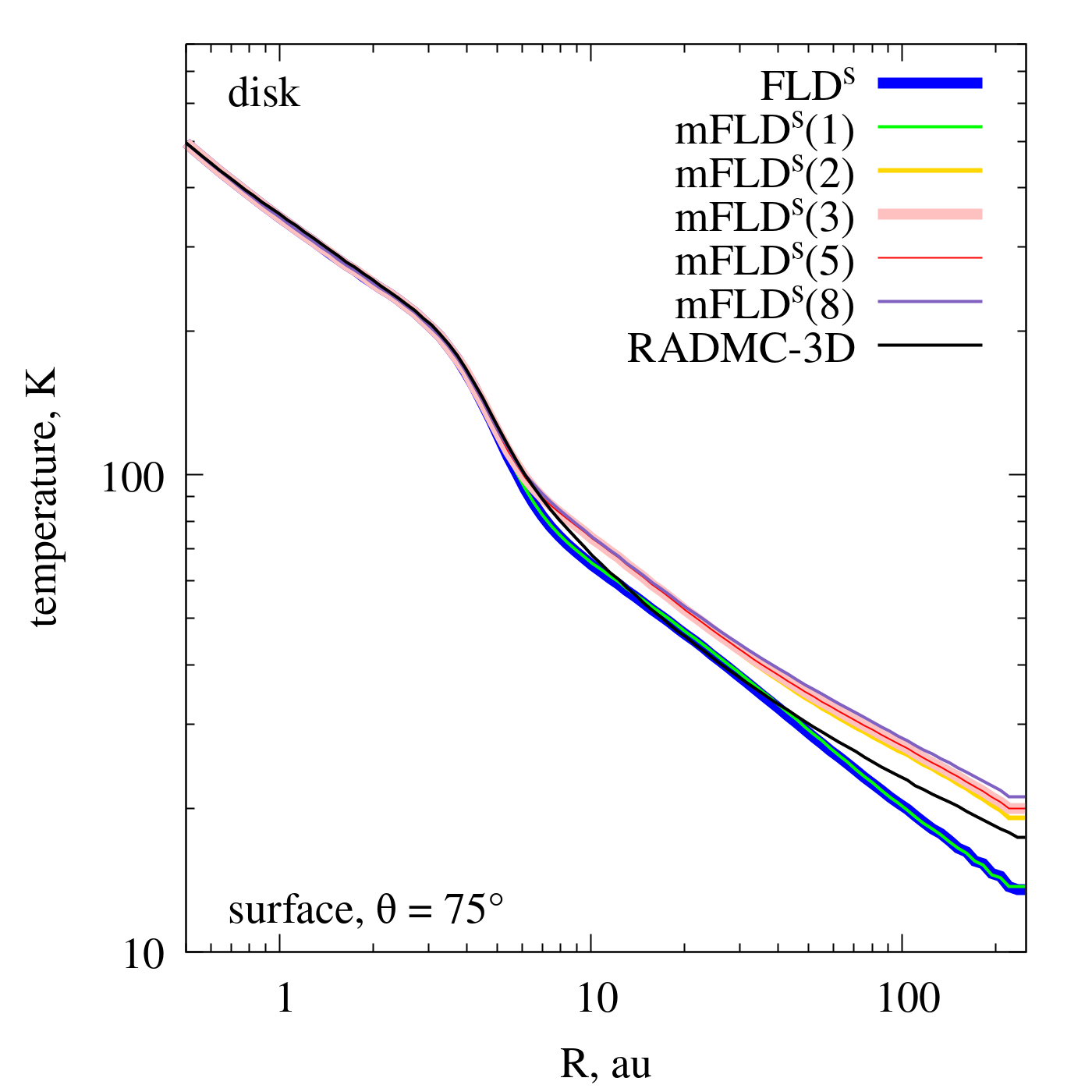}
\includegraphics[angle=0,width=0.48\textwidth]{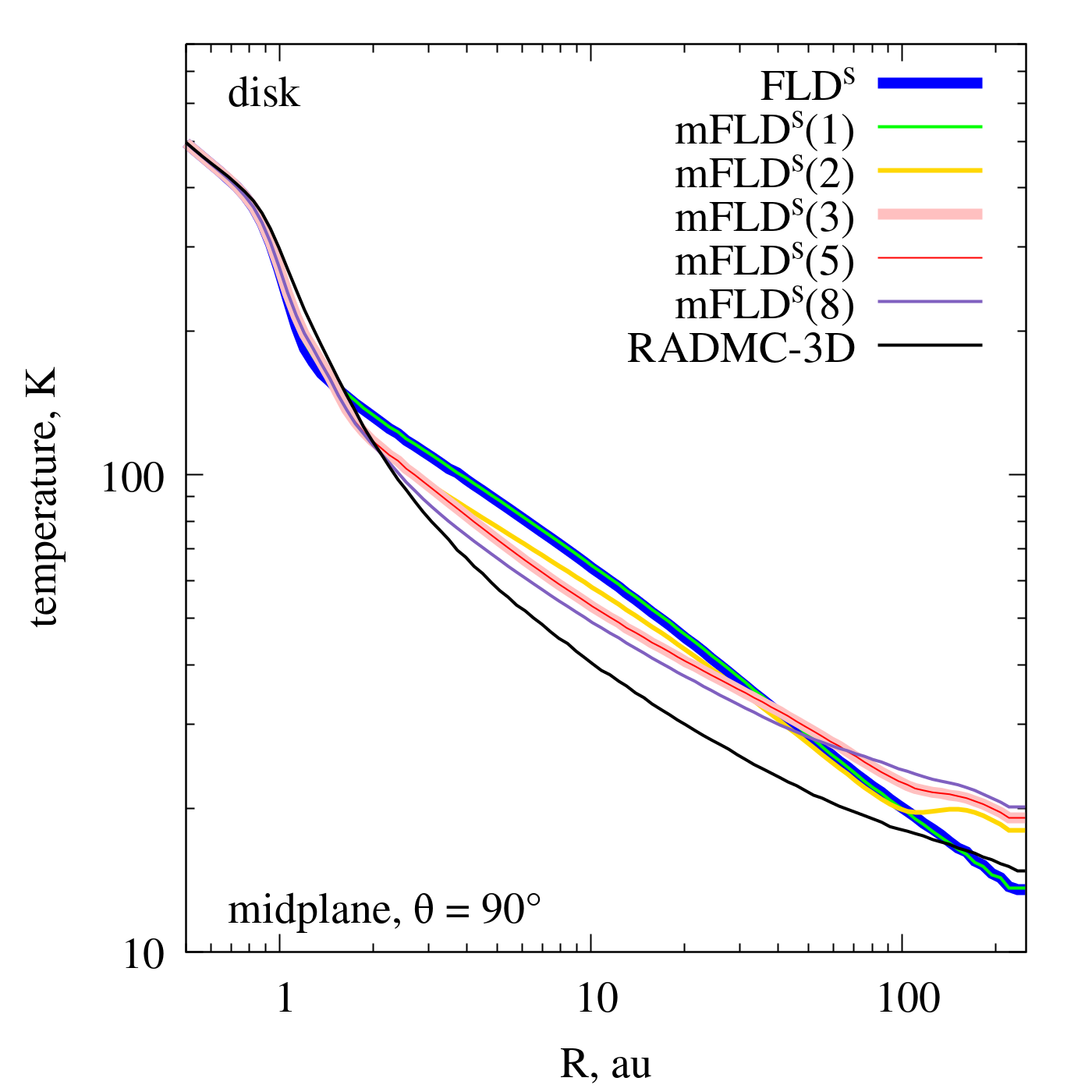}
\caption{Upper panel: scheme of dividing the frequency range into
channels. Lower panels: stationary temperature distributions for the
gas-dust disk model, obtained by the mFLD$^{\rm s}$ method with a
different number of frequency channels.}
\label{fig:compareA8}
\end{figure*}

Figure~\ref{fig:compareA8} shows the temperature distributions for the
protoplanetary disk model, calculated by the mFLD$^{\rm s}$ method with
different divisions of the frequency range into channels. We performed
calculations using 1, 2, 3, 5, and 8 channels, the boundaries of which
are indicated on the upper panel of Fig.~\ref{fig:compareA8}. The number
of channels used in the mFLD$^{\rm s}$ method will henceforth be
indicated in parentheses after it. Recall that the previously described
results correspond to the mFLD$^{\rm s}$(5) method.

As expected, the results of the mFLD$^{\rm s}$(1) method coincide with
the results of FLD$^{\rm s}$. In general, the most significant changes in
the results are observed when moving from mFLD$^{\rm s}$(1) to mFLD$^{\rm
s}$(2) -- the midplane temperature distribution of the disk ceases to
obey the law $T\propto R^{-1/2}$ and in morphology becomes similar to the
reference (obtained by RADMC-3D) distribution. With the used divisions
into frequency ranges, the results of the mFLD$^{\rm s}$(3) and
mFLD$^{\rm s}$(5) methods are already almost identical. Increasing the
number of frequency channels to 8 does not lead to significant changes
--- the mFLD$^{\rm s}$(8) method gives a slightly lower temperature
inside 70 au and a slightly higher temperature beyond 70 au compared to
the mFLD$^{\rm s}$(5) method.

\begin{table}
\centering
\caption{Performance of the
FLD$^{\rm s}$ and mFLD$^{\rm s}$ Methods}
\begin{tabular}{c|c}
\hline
Method & Calculation Time \\
\hline
FLD$^{\rm s}$     & 4 min \\
mFLD$^{\rm s}$(1) & 107 min \\
mFLD$^{\rm s}$(2) & 79 min \\
mFLD$^{\rm s}$(3) & 66 min \\
mFLD$^{\rm s}$(5) & 164 min \\
mFLD$^{\rm s}$(8) & 353 min \\
\hline
\end{tabular}
\label{tab_models}
\end{table}

Table \ref{tab_models} shows the calculation times of the noted methods
in the problem of establishing a stationary temperature distribution. All
calculations were performed on a grid of $100\times50$ cells, starting
from the moment $10^{-4}$ years to the moment $10^{4}$ years, with a
variable time step. A total of 243 time steps were performed. All
simulations were performed on a laptop with an AMD Ryzen~5 5500U processor
using the Intel Fortran compiler ver. 2021.1.2 (without additional options) and the
Ubuntu~22.04 operation system. In the current implementation, the mFLD$^{\rm
s}$ method turned out to be significantly slower than the FLD$^{\rm s}$
method, which makes its use in the current program implementation in
hydrodynamic calculations difficult. This is largely due to the fact that
the algorithm of the method has not yet been optimized by us --- this is
the goal of future work. We also note the non-monotonic change in
calculation time with the number of spectral ranges. This behavior is a
consequence of the different convergence history of Newton iterations. In
particular, in the case of mFLD$^{\rm s}$(1), the largest number of
iterations was required for convergence. However, we note that even in
the current implementation, the mFLD$^{\rm s}$ method can be used to
solve a number of non-stationary problems. In the future, we plan to use
it to study the effect of luminosity bursts on the heating of a protoplanetary disk.

\section{Conclusion}
In this work, we continued the analysis of the model for calculating the
thermal structure of an axisymmetric protoplanetary disk, initiated in
the paper by~\cite{2024ARep...68.1045P}. The thermal model is based on the
well-known Flux-Limited Diffusion (FLD) approximation with a separate
calculation of heating by direct stellar radiation (the FLD$^{\rm s}$
method). In addition to the previously described FLD$^{\rm s}$ model with
spectrum-averaged opacities, this work presents the multiband
model mFLD$^{\rm s}$, in which the spectrum of thermal radiation is
divided into several frequency bands. The implicit finite-difference
scheme for the equations of thermal radiation diffusion is reduced to a
system of linear algebraic equations, for the solution of which a
modified Gauss method for inverting the sparse hypermatrix of the
original system is proposed. We present the results of
testing the described methods, as well as the results of modeling the
thermal structure of a protoplanetary disk. The results of testing the
FLD$^{\rm s}$ and mFLD$^{\rm s}$ methods allow us to assert their correct
implementation for modeling radiation transfer in an axisymmetric
approximation in a spherical coordinate system. The main conclusions of
the analysis of the FLD$^{\rm s}$ and mFLD$^{\rm s}$ methods:
\begin{enumerate}
\item The mFLD$^{\rm s}$ model allowed us to improve the agreement with
the reference distribution of the stationary temperature in the disk
compared to the FLD$^{\rm s}$ model. In particular, the radial
temperature profile from mFLD$^{\rm s}$ in the disk midplane has a variable
slope in accordance with the results of calculations
by the Monte Carlo method implemented in the RADMC-3D code. The mFLD$^{\rm s}$ model
also qualitatively reproduces the non-isothermality of the temperature
distribution along the vertical direction near the midplane,
which is not provided by the FLD$^{\rm s}$ method. However, quantitative
differences remain between the reference temperature values and the
results of mFLD$^{\rm s}$. These differences are likely due to the
diffusive nature of the FLD approximation and, in particular, the
isotropy of the diffusion coefficients (flux limiters).
\item The characteristic timescales for the disk to reach thermal equilibrium
in the test problem of heating the disk from an initial level of 2.73 K
within the mFLD$^{\rm s}$ model turned out to be significantly shorter
than in the FLD$^{\rm s}$ model. This is a consequence of the fact that
in the multiband model, radiation can more easily redistribute through
the disk. The possible difference in thermal timescales between
the single frequency channel and multi-channel approximations should be taken into account
when modeling non-stationary processes in protoplanetary disks within
FLD-based models.
\item The study of the accuracy of mFLD$^{\rm s}$ depending on the number
of spectral channels shows that the presence of only two to three
channels significantly improves the results compared to FLD$^{\rm s}$.
\end{enumerate}

\section{Acknowledgments}
The authors are grateful to the reviewer for suggestions to improve the article. The research was carried out with the support of the Russian Science
Foundation grant No. 22-72-10029, https://rscf.ru/project/22-72-10029/.

\bibliographystyle{mnras}
\bibliography{mflds}

\begin{thebibliography}{}
\makeatletter
\relax
\def\mn@urlcharsother{\let\do\@makeother \do\$\do\&\do\#\do\^\do\_\do\%\do\~}
\def\mn@doi{\begingroup\mn@urlcharsother \@ifnextchar [ {\mn@doi@}
  {\mn@doi@[]}}
\def\mn@doi@[#1]#2{\def\@tempa{#1}\ifx\@tempa\@empty \href
  {http://dx.doi.org/#2} {doi:#2}\else \href {http://dx.doi.org/#2} {#1}\fi
  \endgroup}
\def\mn@eprint#1#2{\mn@eprint@#1:#2::\@nil}
\def\mn@eprint@arXiv#1{\href {http://arxiv.org/abs/#1} {{\tt arXiv:#1}}}
\def\mn@eprint@dblp#1{\href {http://dblp.uni-trier.de/rec/bibtex/#1.xml}
  {dblp:#1}}
\def\mn@eprint@#1:#2:#3:#4\@nil{\def\@tempa {#1}\def\@tempb {#2}\def\@tempc
  {#3}\ifx \@tempc \@empty \let \@tempc \@tempb \let \@tempb \@tempa \fi \ifx
  \@tempb \@empty \def\@tempb {arXiv}\fi \@ifundefined
  {mn@eprint@\@tempb}{\@tempb:\@tempc}{\expandafter \expandafter \csname
  mn@eprint@\@tempb\endcsname \expandafter{\@tempc}}}

\bibitem[\protect\citeauthoryear{{Armitage}}{{Armitage}}{2015}]{2015arXiv150906382A}
{Armitage} P.~J.,  2015, \mn@doi [arXiv e-prints] {10.48550/arXiv.1509.06382},
  \href {https://ui.adsabs.harvard.edu/abs/2015arXiv150906382A} {p.
  arXiv:1509.06382}

\bibitem[\protect\citeauthoryear{{Dullemond}}{{Dullemond}}{2002}]{2002A&A...395..853D}
{Dullemond} C.~P.,  2002, \mn@doi [\aap] {10.1051/0004-6361:20021300}, \href
  {https://ui.adsabs.harvard.edu/abs/2002A&A...395..853D} {395, 853}

\bibitem[\protect\citeauthoryear{{Dullemond}, {Juhasz}, {Pohl}, {Sereshti},
  {Shetty}, {Peters}, {Commercon}  \& {Flock}}{{Dullemond}
  et~al.}{2012}]{2012ascl.soft02015D}
{Dullemond} C.~P.,  {Juhasz} A.,  {Pohl} A.,  {Sereshti} F.,  {Shetty} R.,
  {Peters} T.,  {Commercon} B.,   {Flock} M.,  2012, {RADMC-3D: A multi-purpose
  radiative transfer tool}, Astrophysics Source Code Library, record
  ascl:1202.015

\bibitem[\protect\citeauthoryear{{Lesur} et~al.,}{{Lesur}
  et~al.}{2023}]{2023ASPC..534..465L}
{Lesur} G.,  et~al., 2023, in {Inutsuka} S.,  {Aikawa} Y.,  {Muto} T.,
  {Tomida} K.,   {Tamura} M.,  eds,  Astronomical Society of the Pacific
  Conference Series Vol. 534, Protostars and Planets VII. p.~465

\bibitem[\protect\citeauthoryear{{Levermore} \& {Pomraning}}{{Levermore} \&
  {Pomraning}}{1981}]{1981ApJ...248..321L}
{Levermore} C.~D.,  {Pomraning} G.~C.,  1981, \mn@doi [\apj] {10.1086/159157},
  \href {https://ui.adsabs.harvard.edu/abs/1981ApJ...248..321L} {248, 321}

\bibitem[\protect\citeauthoryear{{Mihalas}}{{Mihalas}}{1978}]{1978stat.book.....M}
{Mihalas} D.,  1978, {Stellar atmospheres}

\bibitem[\protect\citeauthoryear{{Pavlyuchenkov}}{{Pavlyuchenkov}}{2024}]{2024ARep...68.1045P}
{Pavlyuchenkov} Y.~N.,  2024, \mn@doi [Astronomy Reports]
  {10.1134/S106377292470094X}, \href
  {https://ui.adsabs.harvard.edu/abs/2024ARep...68.1045P} {68, 1045}

\bibitem[\protect\citeauthoryear{{Pavlyuchenkov}, {Wiebe}, {Akimkin},
  {Khramtsova}  \& {Henning}}{{Pavlyuchenkov}
  et~al.}{2012}]{2012MNRAS.421.2430P}
{Pavlyuchenkov} Y.~N.,  {Wiebe} D.~S.,  {Akimkin} V.~V.,  {Khramtsova} M.~S.,
  {Henning} T.,  2012, \mn@doi [\mnras] {10.1111/j.1365-2966.2012.20480.x},
  \href {https://ui.adsabs.harvard.edu/abs/2012MNRAS.421.2430P} {421, 2430}

\bibitem[\protect\citeauthoryear{{Teyssier} \& {Commer{\c{c}}on}}{{Teyssier} \&
  {Commer{\c{c}}on}}{2019}]{2019FrASS...6...51T}
{Teyssier} R.,  {Commer{\c{c}}on} B.,  2019, \mn@doi [Frontiers in Astronomy
  and Space Sciences] {10.3389/fspas.2019.00051}, \href
  {https://ui.adsabs.harvard.edu/abs/2019FrASS...6...51T} {6, 51}

\bibitem[\protect\citeauthoryear{{Vaytet}, {Audit}, {Chabrier},
  {Commer{\c{c}}on}  \& {Masson}}{{Vaytet} et~al.}{2012}]{2012A&A...543A..60V}
{Vaytet} N.,  {Audit} E.,  {Chabrier} G.,  {Commer{\c{c}}on} B.,   {Masson} J.,
   2012, \mn@doi [\aap] {10.1051/0004-6361/201219427}, \href
  {https://ui.adsabs.harvard.edu/abs/2012A&A...543A..60V} {543, A60}

\bibitem[\protect\citeauthoryear{{W{\"u}nsch}}{{W{\"u}nsch}}{2024}]{2024FrASS..1146812W}
{W{\"u}nsch} R.,  2024, \mn@doi [Frontiers in Astronomy and Space Sciences]
  {10.3389/fspas.2024.1346812}, \href
  {https://ui.adsabs.harvard.edu/abs/2024FrASS..1146812W} {11, 1346812}

\bibitem[\protect\citeauthoryear{{van der Holst} et~al.,}{{van der Holst}
  et~al.}{2011}]{2011ApJS..194...23V}
{van der Holst} B.,  et~al., 2011, \mn@doi [\apjs]
  {10.1088/0067-0049/194/2/23}, \href
  {https://ui.adsabs.harvard.edu/abs/2011ApJS..194...23V} {194, 23}

\makeatother
\end{thebibliography}

\end{document}